\documentclass[5p]{elsarticle}

\usepackage{amsmath}
\usepackage{amsfonts}
\usepackage{amssymb}
\usepackage{enumerate}
\usepackage{verbatim}
\usepackage{csquotes}
\usepackage{hyphenat}

\usepackage{hyperref}
\usepackage{color}
\definecolor{lightgray}{gray}{0.5}

\newcommand{\he}{$^4\mathrm{He}$ }
\newcommand{\hen}{$^4\mathrm{He}$}
\newcommand{\hee}{$^3\mathrm{He}$ }

\definecolor{orange}{rgb}{1,0.5,0}

\newcommand{\spider}{{\sc Spider} }
\newcommand{\spidern}{{\sc Spider}}

\newcommand{\bicep}{{\sc BICEP1 }}

\newcommand{\biceptwo}{{\sc BICEP2 }}
\newcommand{\biceptwon}{{\sc BICEP2}}

\newcommand{\mrm}[1]{\mathrm{#1}}

\title{The Thermal Design, Characterization, \\and Performance of the \spider Long-Duration Balloon Cryostat}

\fauthor[a]{J.E.~Gudmundsson}
\author[]{for the \spider Collaboration:}
\author[b]{P.A.R.~Ade}
\author[c]{M.~Amiri}
\author[a,d]{S.J.~Benton}
\author[f,g]{J.J.~Bock}
\author[h]{J.R.~Bond}
\author[m]{S.A.~Bryan}
\author[n]{H.C.~Chiang}
\author[i]{C.R.~Contaldi}
\author[f,g]{B.P.~Crill}
\author[f,g]{O.~Dore}
\author[p]{J.P.~Filippini}
\author[a]{A.A.~Fraisse}
\author[a]{A. Gambrel}
\author[d]{N.N.~Gandilo}
\author[q,c]{M.~Hasselfield}
\author[c]{M.~Halpern}
\author[j]{G.~Hilton}
\author[f]{W.~Holmes}
\author[g]{V.V.~Hristov}
\author[k]{K.D.~Irwin}
\author[a]{W.C.~Jones}
\author[a]{Z.~Kermish}
\author[l]{C.J.~MacTavish}
\author[g]{P.V.~Mason}
\author[f]{K.~Megerian}
\author[g]{L.~Moncelsi}
\author[e]{T.E.~Montroy}
\author[g]{T.A.~Morford}
\author[e]{J.M.~Nagy}
\author[d,o,s]{C.B.~Netterfield}
\author[a]{A.S.~Rahlin}
\author[j]{C.D.~Reintsema}
\author[e]{J.E.~Ruhl}
\author[g]{M.C.~Runyan}
\author[o]{J.A.~Shariff}
\author[r,o]{J.D.~Soler}
\author[g,f]{A.~Trangsrud}
\author[b]{C.~Tucker}
\author[g]{R.S.~Tucker}
\author[f]{A.D.~Turner}
\author[c]{D.V.~Wiebe}
\author[a]{E. Young}
\address[a]{Department of Physics, Princeton University, Princeton, NJ, U.S.A.}
\address[b]{School of Physics and Astronomy, Cardiff University, Cardiff,
U.K.}
\address[c]{Department of Physics and Astronomy, University of British
Columbia, Vancouver, BC, Canada}
\address[d]{Department of Physics, University of Toronto, Toronto, ON,
Canada}
\address[e]{Department of Physics, Case Western Reserve University,
Cleveland, OH, U.S.A.}
\address[f]{Jet Propulsion Laboratory, Pasadena, CA, U.S.A.}
\address[g]{Division of Physics, Mathematics \& Astronomy, California Institute of Technology,
Pasadena, CA, U.S.A.}
\address[h]{Canadian Institute for Theoretical Astrophysics, University
of Toronto, Toronto, ON, Canada}
\address[i]{Theoretical Physics, Blackett Laboratory, Imperial College, London, U.K.}
\address[j]{National Institute of Standards and Technology, Boulder, CO, U.S.A.}
\address[k]{Department of Physics, Stanford University, Stanford, CA, U.S.A.}
\address[l]{Kavli Institute for Cosmology, University of Cambridge, Cambridge, U.K.}
\address[m]{Arizona State University, Tempe, AZ, U.S.A.}
\address[n]{School of Mathematics, Statistics \& Computer Science, University of KwaZulu-Natal, Durban, South Africa}
\address[o]{Department of Astronomy and Astrophysics, University of Toronto, Toronto, ON, Canada}
\address[p]{Department of Physics, University of Illinois at Urbana-Champaign, IL, U.S.A.}
\address[q]{Department of Astrophysical Sciences, Princeton University, Princeton, NJ, U.S.A.}
\address[r]{Institut d'Astrophysique Spatiale, Orsay, France}
\address[s]{Canadian Institute for Advanced Research CIFAR Program in Cosmology and
Gravity, Toronto, ON, Canada}

\begin{document}

\begin{abstract}
We describe the \spider flight cryostat, which is designed to cool
six millimeter-wavelength telescopes during an Antarctic
long-duration balloon flight. The cryostat, one of the largest to have
flown on a stratospheric payload, uses liquid \he to deliver cooling
power to stages at 4.2 and 1.6~K. Stainless steel capillaries facilitate a
high flow impedance connection between the main liquid helium tank and
a smaller superfluid tank, allowing the latter to operate at 1.6~K as
long as there is liquid in the 4.2~K main tank. Each telescope houses a
closed cycle \hee adsorption refrigerator that further cools the focal
planes down to 300~mK. Liquid helium vapor from the main tank is
routed through heat exchangers that cool radiation shields, providing
negative thermal feedback. The system performed
successfully during a 17~day flight in the 2014--2015 Antarctic summer. The
cryostat had a total hold time of 16.8~days, with 15.9~days occurring during flight.
\end{abstract}

\maketitle

\section{Introduction}

Numerous experiments have characterized the spectrum, morphology and,
most recently, the polarization of the Cosmic Microwave Background
(CMB)~\cite{Polarbear2014,BICEP2_I,ACT_Naess2014,Crites2014,Planck2015_Cosmology}.
Many of these experiments rely on cryogenically-cooled receivers 
to suppress thermal noise that would otherwise dominate the CMB
signal. \spider is a microwave polarimeter employing cryogenic
bolometers sensitive to broad band radiation at 95 and 150 GHz,
that is designed to image the CMB polarization
with the aim of constraining models of the early universe
\cite{Fraisse2011,Rahlin2014}.

\spider deploys multiple arrays of superconducting transition edge
sensors on a stratospheric balloon platform supplied by NASA\rq s
Columbia Scientific Balloon Facility (CSBF). The
design requirements include the cooling of six telescopes,
representing a total aperture of nearly 0.5 square meters, to
300~mK for approximately 20 days while providing continuous cooling power to
auxiliary temperature stages at 1.6, 4.2, 30, and 150 K, and staying within
the mass and power constraints set by the balloon platform. A
helium-only system that employs two cryogenic volumes, a 1284~L main
tank and a 16~L superfluid tank, meets these requirements. 
The superfluid tank is continuously
supplied via capillaries that connect it to the main tank while
maintaining a one-atmosphere pressure differential between the two
tanks \cite{Delong1970}. Vapor cooled aluminum shells surround the 
two tanks and provide radiation shielding and negative thermal feedback.

\begin{figure*}[t!]
\begin{center}
\begin{tabular}{c}
\includegraphics[width = 0.99\textwidth]{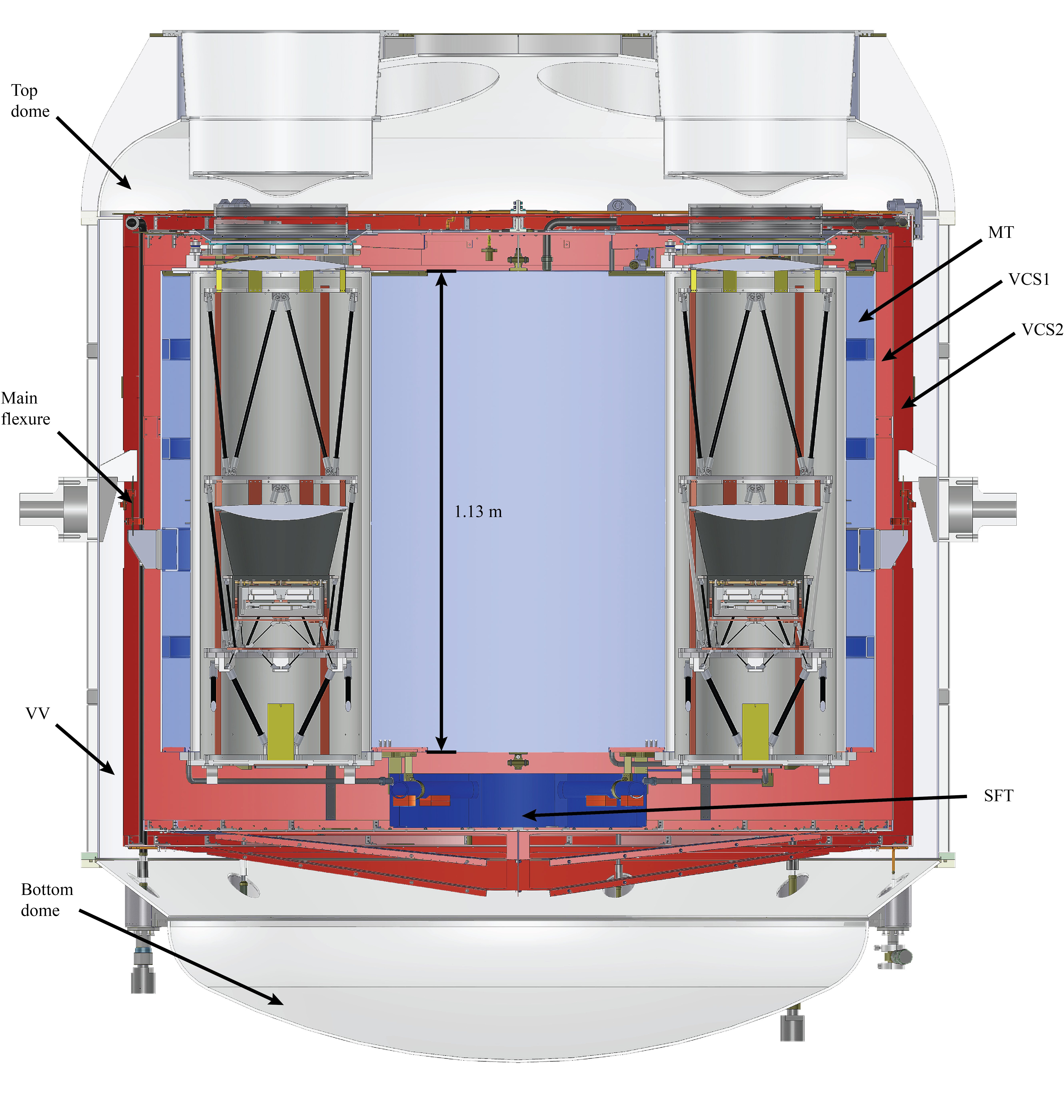}
\end{tabular}
\end{center}
    \caption[A cross-sectional of the flight cryostat]{A cross section
      rendering of the flight cryostat with two telescopes
      visible. The outermost layer, the vacuum vessel, surrounds the
      two vapor cooled shields, shown in dark (VCS2) and light (VCS1) red. At the
      center, the main and superfluid tanks are colored in light and
      dark blue, respectively. Five plumbing lines exit at the bottom:
      the main and superfluid tank fill and vent lines, and the VCS
      vent line.}
\label{fig:theo_cad}
\end{figure*}

The construction of the \spider flight
cryostat was preceded by the initial fabrication of three smaller test
cryostats.  Each of these test cryostats accommodated a single telescope,
and helped verify both the design and manufacturing
methods applied to the larger flight cryostat.  Thermal qualification
of those systems showed that a helium-only cryostat with two vapor
cooled shields could successfully cool an entire \spider telescope
while maintaining minimal in-band loading from warmer temperature
stages, and a reasonable cryogen boil-off rate. Additionally, the test
cryostats demonstrated the performance of: 1) the heat exchangers
cooling intermediate stages; 2) multilayer insulation and its
distribution across radiation shields; and 3) thermal conduction paths
to sub-kelvin stages, critical for the operation of adsorption
refrigerators. After qualification, one of the three \spider cryostats
was provided to the \biceptwo instrument, maintaining that system at
cryogenic temperatures continuously for three years at the South
Pole~\cite{BICEP2_II}.  Based on the performance of these test
cryostats, the decision was made to go ahead with the construction of the
\spider flight cryostat.

The construction of the \spider flight cryostat, by
Redstone Aerospace, began in the fall of 2008 and was 
completed in the summer of 2009.\footnote{\textit{Redstone Aerospace}, Longmont, CO.}
After preliminary testing at Redstone, the cryostat was transported to
Princeton in January 2010 where numerous cryogenic runs were
conducted. A description of the cryostat along with an early
performance evaluation can be found in \cite{Gudmundsson2010}. In
2012, the vacuum integrity of the flight cryostat was tested in the
Spacecraft Propulsion Research Facility at the NASA Glenn Research
Center in Sandusky, Ohio. This allowed us to verify the performance of 
numerous hermetic seals under flight-like pressure conditions.
Integration and testing of the entire balloon
payload was performed at CSBF in Palestine, Texas, in the
summer of 2013, with the aim of deploying the payload later that
year. The shutdown of the United States federal government in the fall
of 2013 caused a one-year delay in our Antarctic operations.

The \spider flight cryostat was launched on a long-duration balloon
(LDB) from McMurdo station in Antarctica on January 1, 2015. The
cryostat ran out of liquid helium 16 days after its last fill. The
flight was terminated 16.5~days after launch, as the payload was
warming up and beginning to drift out to sea.

\section{Design}
\subsection{Volume, Mass, and General Cryogenic Layout}

The cylindrical flight cryostat has five main components that
are illustrated in Figure~\ref{fig:theo_cad}. Starting from the
inside, the components are named: Superfluid Tank (SFT), Main Tank
(MT), Vapor Cooled Shields 1 and 2 (VCS1, VCS2), and Vacuum Vessel
(VV). The bulk of the cryostat is made of aluminum 1100 (VCS1, VCS2),
chosen for its high thermal conductivity, and aluminum 5083 (VV, MT, SFT), 
that, although a poor thermal conductor, maintains its strength after welding. 
The cryogenic assembly consists of a 1284 L liquid helium (LHe) main tank, 
in the shape of a cylinder penetrated completely by seven cylindrical holes.  
This shape allows each telescope to be surrounded by a uniform 
4~K environment.  The main tank is connected by a capillary 
system to a 16 L superfluid tank (see Sections \ref{sec:sft} and \ref{sec:capassy}). 
The steady state
parasitic load of the instruments and open apertures results in a base
mass flow rate of helium that obviates the need for a liquid nitrogen
guard; the enthalpy of the helium vapor is sufficient to provide the
required cooling power above 80~K.  VCS1 and VCS2 surround both
tanks and serve as radiation shields, while intercepting conduction
and accommodating filters, which need to be maintained at low
temperatures to reduce in-band parasitics (see Section
\ref{sec:filters}).

The lightweight design of the main tank, superfluid tank, and the
vacuum vessel, each with an average wall thickness of approximately 4
mm, allows such a large system to be deployed on a long-duration
balloon. The whole cryogenic assembly, excluding telescope
parts, cryogens, and cabling, weighs 850~kg. In total, the entire
\spider payload weighed approximately 2810~kg, not including the
support instrumentation package and ballast, both supplied by CSBF.

\begin{figure}[t!]
\begin{center}
\begin{tabular}{c}
\includegraphics*[angle=0,width=0.46\textwidth]{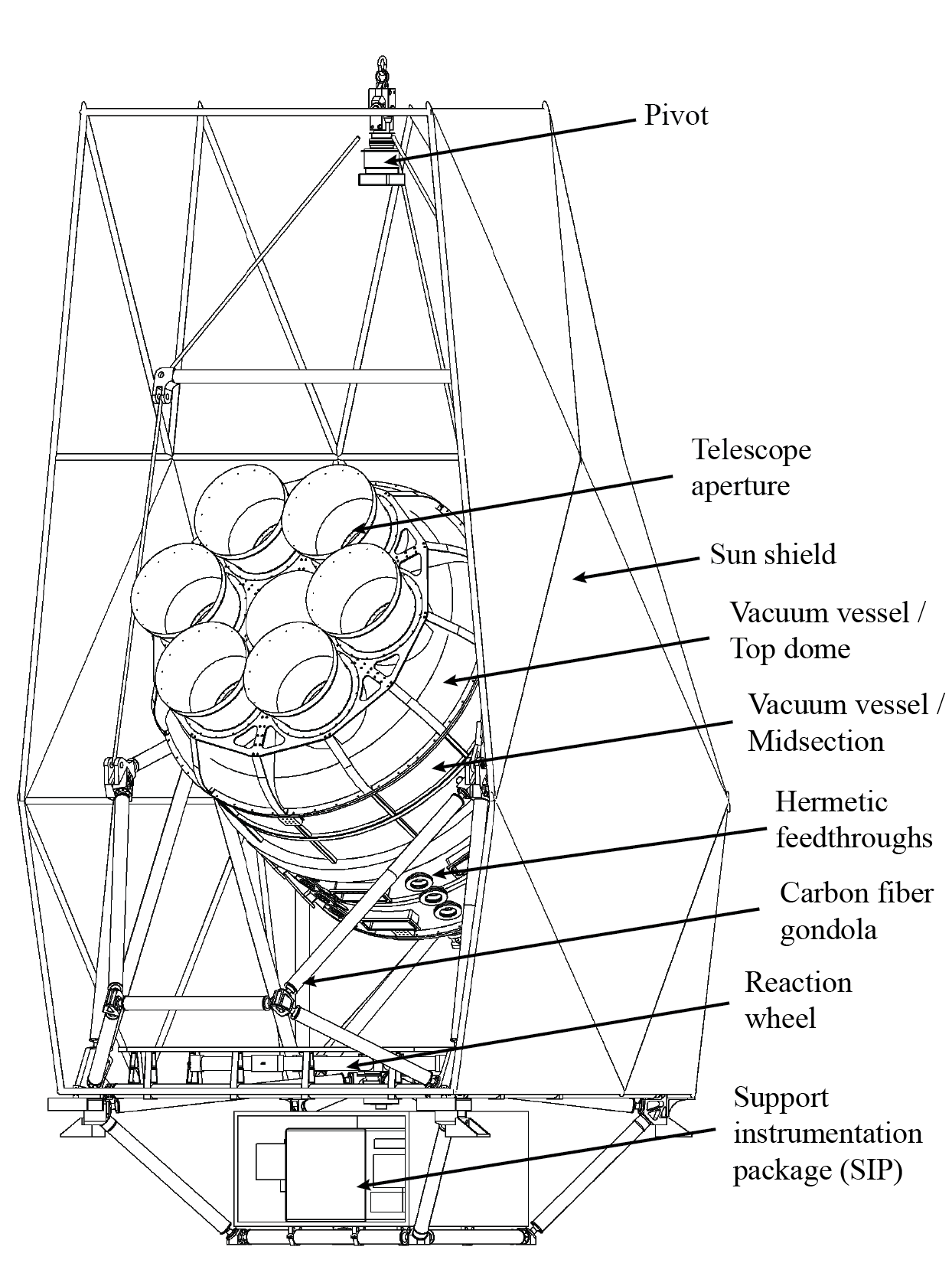}
\end{tabular}
\caption[CAD model of the \spider payload]{A simplified CAD model of
  the \spider payload, including the outer vacuum vessel (VV), gondola,
  sun-shields, and CSBF Support Instrumentation Package (SIP). Ports
  on the front side of the cylinder provide hermetic connections to
  housekeeping electronics (not shown). The sun-shields are composed
  of a lightweight carbon fiber frame which is tiled with
  foam and aluminized Mylar. The support instrumentation package houses antenna power and
  control systems, and is mounted underneath the cryostat at the
  bottom of the payload. This figure omits a few key components of the
  payload for increased clarity; for example, it does not include the
  solar panels that were mounted on the port side of the sun-shields.}
\label{fig:spider}
\end{center}
\end{figure}

\begin{figure}[b!]
\begin{center}
\begin{tabular}{c}
\includegraphics*[angle=0,width=0.49\textwidth]{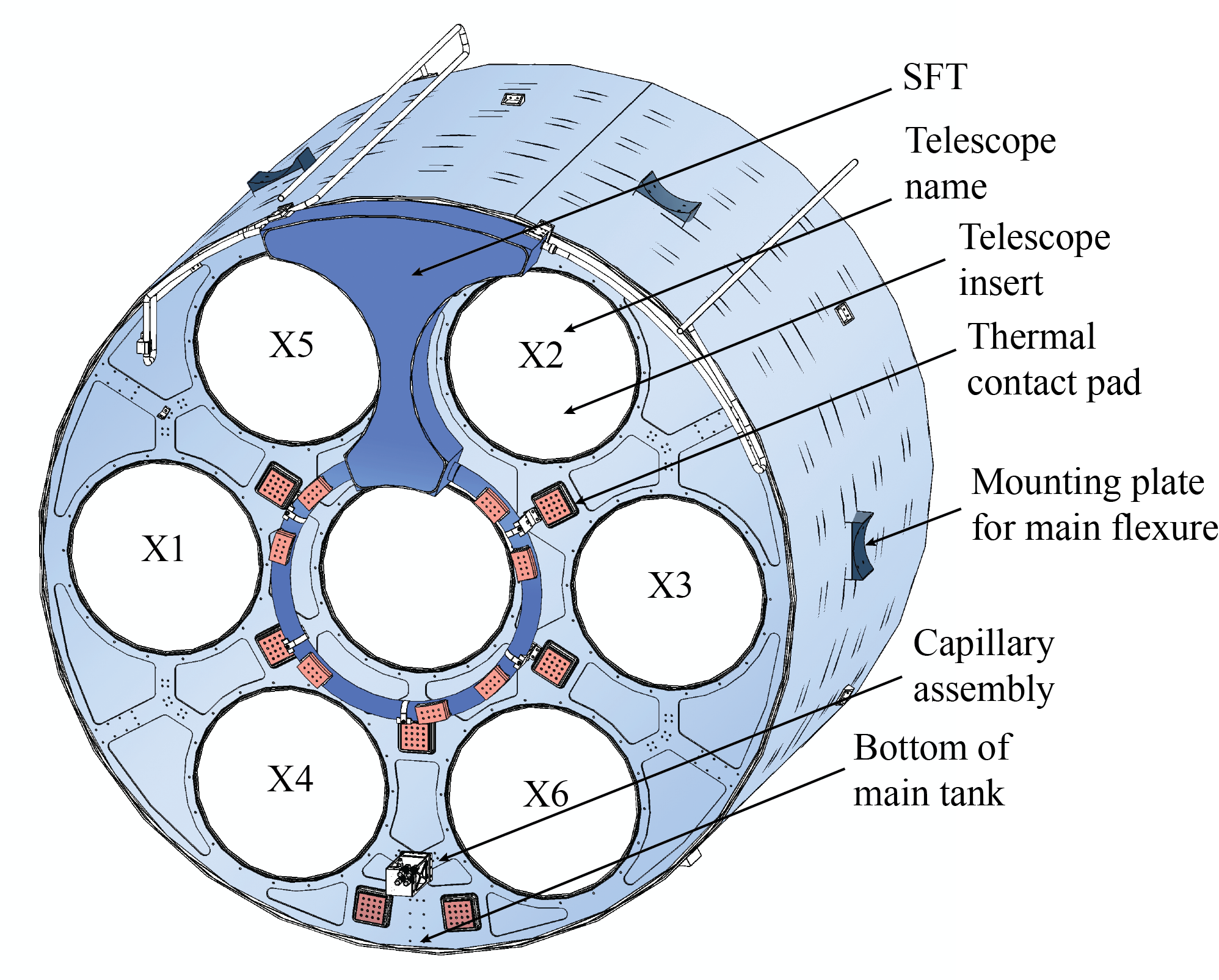}
\end{tabular}
\caption[CAD model of the vacuum vessel]{CAD model of the main tank
  and superfluid tank assemblies as viewed from the bottom. 
  Note the seven extruded cuts through the main tank, which house 
  the six telescope inserts and a cabling feed-through in the center.
  The SFT has a ring-like structure that connects to another
  larger cryogenic volume located under the MT. There are seven
  explosion-bonded contact pads on both the MT and SFT. These provide
  the 4~K and 1.6~K thermal anchor points for each telescope. The
  copper straps used for this purpose are not shown on this
  schematic. The cylindrical MT is $1.69 \:\mrm{m}$ in diameter, $1.13
  \:\mrm{m}$ long, and weighs $240 \:\mrm{kg}$. The insert diameter is
  $419 \:\mrm{mm}$ and the thickness of the MT walls varies between
  $4$ and $6 \:\mrm{mm}$. Observations are performed with the cryostat
  at a range of elevations from 22--49~degrees, such that the bulk of the SFT is
  above the ring which holds only about $0.5 \:\mrm{L}$. The
  capillary assembly can be seen at the bottom of the figure, between
  two thermal contact pads. The telescope inserts are labeled with
  the telescope names, X1--X6, that populated them during flight. The
  bellows connecting the capillary assembly to both the MT and SFT are
  omitted from this figure.}
\label{fig:mtsft}
\end{center}
\end{figure}

\subsection{Gondola and Sun-Shields}

The flight cryostat is mounted on a lightweight gondola structure made
of reinforced carbon fiber that houses the drive mechanisms for 
scanning in both elevation and azimuth (see Figure \ref{fig:spider}). 
Aluminum inserts are glued into hollow carbon fiber-reinforced polymer tubes and then 
bolted into custom-made aluminum joints. The gondola is designed to sustain large angular
acceleration from the initial balloon jerk during launch, as well as a
maximum 10~$g$ vertical acceleration resulting from the parachute 
shock upon termination \cite{Soler2014,Shariff2014}.

\spidern's sun-shields prevent direct illumination of the telescope
baffles by the sun and provide shadowing of the cryostat. The vacuum vessel
 was painted with white Aero\hyp{}glaze A276, providing well
controlled emissivities at infrared and optical wavelengths.\footnote{LORD Corporation, Cary, NC.} Radiative modeling of the integrated payload, that included shielding to help
thermalize the VV shell, suggested that a $\sim$30~K temperature gradient
would develop over the surface of the vacuum vessel at altitude, with the
equilibrium temperature of the upper dome and vacuum windows 
expected to approach --20~$^{\circ}$C. Because 
of low ambient temperatures, a combination of copper gaskets and 
EPDM O-rings were used for hermetic connections on the vacuum vessel.
Additionally, telescope windows were shock tested with liquid nitrogen
to simulate rapid cooling during ascent. 

\subsection{Radiation Shields, Heat Exchangers, and MLI}
Two intermediate vapor cooled radiation shields, VCS1 and VCS2, serve
as thermal sinks for Multi-Layer Insulation (MLI), cabling, and infrared
blocking filters.  These nested shields are mechanically supported from the main tank
and vacuum vessel, respectively. Six compact heat exchangers are
symmetrically placed on the top sides of both VCS1 and VCS2. Helium vapor is 
routed through these flow-restrictive heat
exchangers, cooling the respective stages, and thus providing negative
thermal feedback \cite{Kays, deWitt}. The heat exchangers are made of
stainless steel blocks enclosing densely packed copper mesh (VCS1)
or pellets (VCS2).

Thermal radiative loading on the MT is reduced by the VCSs and MLI. The MLI is a layer construction of 6.4~$\mrm{\mu m}$ thick thermoplastic polymer (Mylar) coated with highly reflective, $35~\mrm{nm}$ thick, aluminum layer separated by 0.1~mm thick spun bound polyester sheets.\footnote{\textit{CAD Cut}, Middlesex, VT.} The compression of the MLI was optimized for minimum pump down time and maximum radiative shielding between the VV and VCS2 and between VCS2 and VCS1. For both, the value is ~14~layers/cm, resulting in 52 and 16~layers of MLI between the VV and VCS2 and between VCS2 and VCS1, respectively. 

\subsection{Flexures, Plumbing, and Thermal Connections}
\label{sec:flexnplumb}

The main tank is supported by the vacuum vessel \linebreak through six
G--10/aluminum flexures symmetrically placed around the cylinder sides (see
Figure~\ref{fig:mtsft}). G--10 is a high yield strength, low thermal conductivity fiberglass 
commonly used for cryogenic flexures \cite{Runyan2008}. 
The flexures are attached with copper straps 
to VCS1 and VCS2 to minimize conducted heat flow 
to the MT \cite{Gudmundsson2010}.

Plumbing lines are made of type 304 stainless steel, due to its low
thermal conductivity and suitability for welding. There are five
plumbing lines leading from the outside of the VV to either the MT or
the SFT. These are: the MT fill and vent lines, which have a $3/4$
inch outer diameter (OD); the SFT fill and vent lines, which have a
$1/2$ inch OD; and the VCS vent line, which has a $1/4$ inch OD. All
plumbing lines have a 0.010~inch wall thickness. Note that the VCS
line represents a high flow-impedance connection from the MT to the
outside.
  
Vent lines are positioned on the cryogenic tanks such that vapor can exit the cryostat when it is tilted through a 22--49~deg elevation range while full of liquid.  All plumbing lines (excluding the VCS vent line itself) are heat sunk at VCS2, but not at VCS1, to prevent thermal shorts between the two shields.  Aluminum-to-stainless steel transitions are made where needed from explosion-bonded blocks that are welded in place.  During flight, the MT fill and vent lines are sealed so that all of the helium vapor from the MT is routed through the VCS heat exchangers.  The length of the MT/SFT vent and fill lines is approximately 2.7~m, while the average travel of gas through the VCS vent line system is about 12~m.  When cooled to 4~K and at equilibrium, we find that approximately 50~torr of differential pressure between the main tank and the VCS outlet will maintain 20~SLPM of flow.

In flight the MT must be maintained at a pressure near one atmosphere so that it does not rapidly evaporate to reach ambient pressure (as the SFT does).  A set of Tavco absolute pressure flow regulators are mounted to the main tank and VCS vent lines to regulate this pressure and prevent any back flow (see Section \ref{sec:launch}).\footnote{\textit{Tavco, Inc.}, Chatsworth, CA, e-mail: \href{mailto:tavcoinc@aol.com}{tavcoinc@aol.com}, fax:
  +1--818--998--8391.}  
  
\begin{figure}[t!]
\begin{center}
\includegraphics[width = 0.45\textwidth]{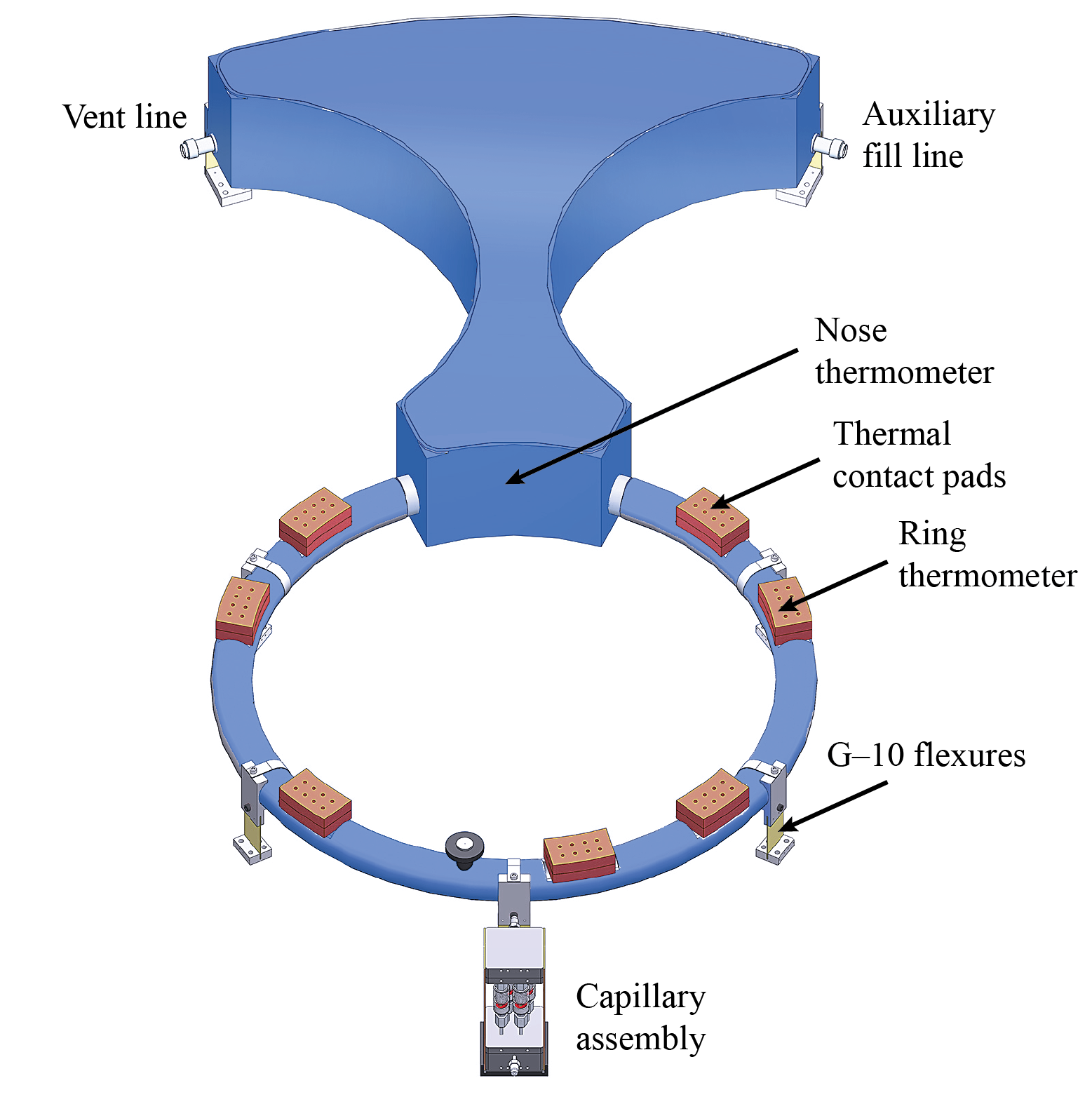}
\end{center}
\caption[The superfluid tank and capillary assembly]{The superfluid
  tank and capillary assembly are mounted to the bottom of the main
  tank (see Figure \ref{fig:mtsft}). The capillary assembly is located
  just below the SFT ring and connected to both tanks using flexible
  1/8~inch bellows tubing. Thermal contact pads are welded into a ring
  that sits below the main volume. The auxiliary fill line and the vent line exit at the
  top to the right and left respectively. The net volume of the
  superfluid tank is 16~L.}
\label{fig:sft_cap}
\end{figure} 

\subsection{Superfluid Tank}
\label{sec:sft}

The SFT is pumped to the ambient pressure of float altitude,
approximately 36 km. This maintains the SFT at about 1.6~K.

Two major requirements motivate the design of the SFT.  The first is conductance: the thermal link between the six science instruments and the SFT must be strong enough to maintain a small temperature gradient when cycling the closed-cycle \hee refrigerators, which results in a 5~mW transient load (per refrigerator) on the 1.6~K cooling stage during this period.  Due to the extremely high thermal conductivity of superfluid \he (ten times that of high-purity copper at the same temperature), the least-mass solution is to use the superfluid itself as the thermal link.  Secondly, the instruments must remain in good contact with the superfluid while \spider scans between 22--49~degrees of elevation, such that the liquid level does not lie in the plane defined by the bottom of the main tank.  Together, these requirements motivate the unusual design shown in Figure~\ref{fig:mtsft}.  A superfluid-filled ring connects to the capillary system and provides excellent thermal conductance to the receivers.  The superfluid surface lies in the larger \enquote{nose} tank at the top of the assembly, ensuring that the ring remains full over the full range of elevation angles.  The instruments are cooled via a series of thermal contact pads around the circumference of the ring.

Some care is needed to make an effective thermal connection between the inserts and the superfluid.  Due to the superfluid’s high conductivity, the thermal impedance between bath and instruments is dominated by Kapitza boundary resistance. To reduce this liquid-to-metal boundary resistance, explosion-bonded aluminum-to-copper transition plates are used to provide thermal contact areas on both the MT and the SFT.\footnote{\textit{High Energy
    Metals, Inc.}, Sequim, WA.} Areas of the aluminum are milled out such that the copper is in direct contact with the liquid cryogen. The transition plates are connected to custom-made copper heat straps which are fabricated by folding 0.005~inch thick high purity copper shim into an 18~layer, 2~inch wide flexible assembly. At an average length of 15~inches, we estimate that the straps conduct approximately 200 mW/K, and find that they supply necessary cooling power to both the 4~K telescope and sub-kelvin cooling stages. The MT thermal contact areas can also be seen in Figure~\ref{fig:mtsft}.

\subsection{Capillary Assembly}
\label{sec:capassy}

The \spider capillary system provides continuous flow of \he from the
main tank to the superfluid tank (see Figure \ref{fig:sft_cap}). The
capillaries are critical for a successful flight because the hold time
of the superfluid tank, when fully charged, is only 4 days, compared
to the targeted 20~day flight duration. The capillaries
provide the superfluid tank with liquid helium as long as there is
liquid in the main tank.

\begin{figure}[b!]
\begin{center}
\begin{tabular}{c}
\includegraphics[width = 0.40\textwidth]{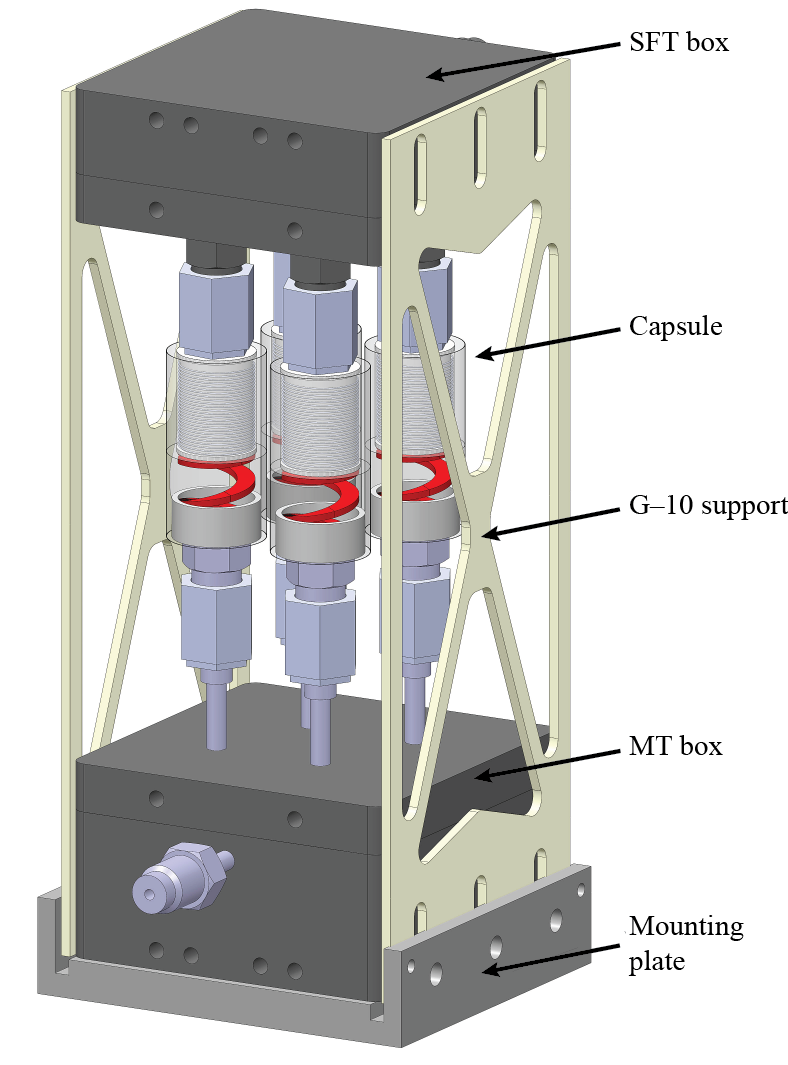}
\end{tabular}
\end{center}
\caption[example]
{The capillary assembly. Four capillaries connect the 4.2~K main tank box (bottom) to the 1.6~K SFT box. The double volume structure is supported by two 1/32~inch thick G--10 flexures. The thermal load conducted through these flexures is negligible compared to the cooling power from the superfluid. Inside the MT box, stainless steel Mott filters, preceding each capillary, prevent ice and other dirt from entering and clogging these high impedance lines. As shown, the assembly is 14.9~cm tall.}
\label{fig:cap}
\end{figure}

The following list describes the main design criteria for the \spider capillary assembly:

\begin{enumerate}
\item Continuous cooling power of approximately 60~mW to counter the steady state loading to the superfluid tank.
\item Base temperature of at most 1.8~K for effective cycling of the six adsorption refrigerators that supply cooling power to the focal planes.
\item No mechanical valves or other moving parts.
\item Robust operations for at least 50 days to accommodate cooldown, characterization, and flight. 
\end{enumerate}

The general design principles of the capillary system are based on 
DeLong et al.\ \cite{Delong1970}, wherein they describe a two
stage \he cryogenic system for a dilution refrigerator. Similar
designs are described in various publications
\cite{Truch2009,Wrubel2011,Das2012,Fujiyoshi1991}.

The steady state loading on the superfluid volume shown in Figure
\ref{fig:sft_cap} has been measured during flight-like
conditions. This measurement, made with the superfluid volume empty
and disconnected from the capillary assembly, suggested a steady state
loading of 40~mW. This measurement is consistent with the measured
hold time of the superfluid tank before the capillaries were
installed. Implementing a 1.5~safety factor we conclude that the
capillary assembly must provide at least 60~mW of continuous cooling
power to the superfluid tank.

Cycling the adsorption refrigerators creates an approximate 5~mW
transient loading on the superfluid tank that lasts for an hour. The
superfluid tank must sustain transients from cycling all six
refrigerators every 24 hours.

\begin{figure}[tb]
\begin{center}
\includegraphics[width = 0.49\textwidth]{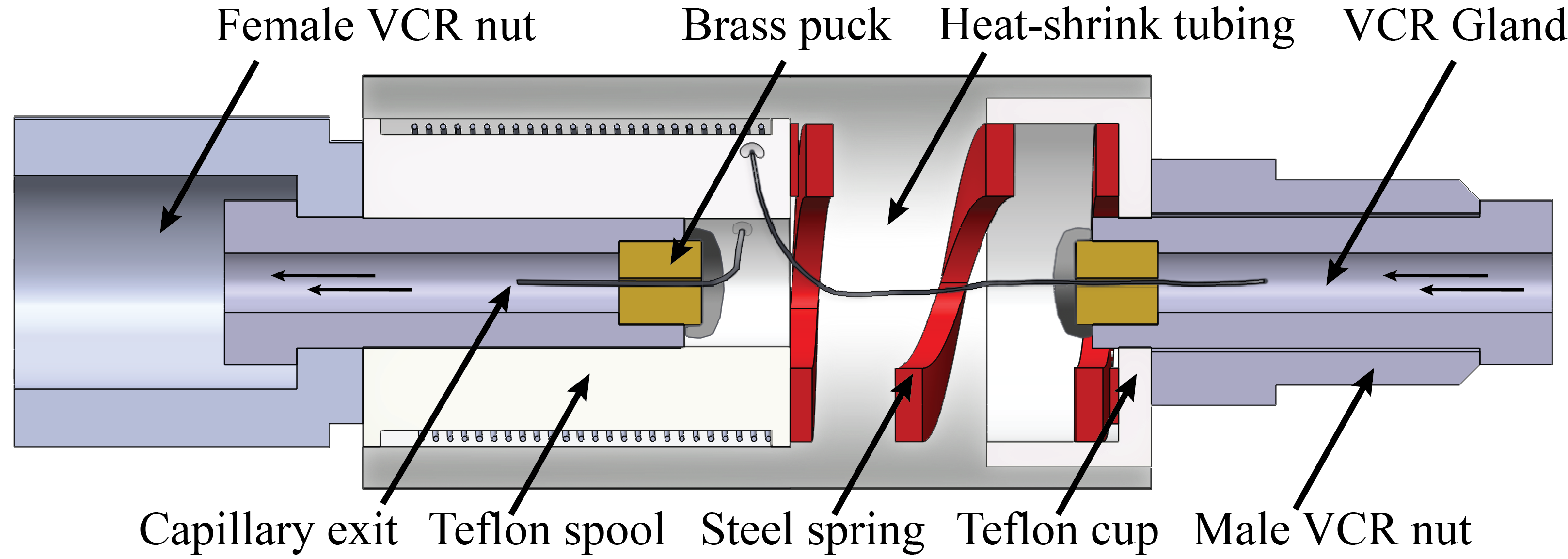}
\end{center}
\caption[A cross section through a capsule]{A cross section
  through one of the capsules, measuring 26.3~mm end-to-end. 
  The ends of the capillaries are located
  inside the Swagelok VCR glands, approximately 1~cm from the brass
  pucks. This way, the ends are shielded from mistreatment. After
  having silver soldered the capillary into the puck on the left, the
  capillary is threaded through a small cavity at the edge of the Teflon
  spool. The spool then slides over the gland, after which the
  capillaries are wrapped around the spool. With only a couple of
  inches remaining, the capillary is threaded through another hole at
  the edge of the spool, then through the Teflon cup, after which the
  other end is silver soldered into the brass puck on the right. The
  stainless steel spring is then carefully wrapped around the
  capillary and secured in the Teflon cup. Finally, heat-shrink tubing
  is positioned around both the Teflon components and spring to shield
  the capillaries and unify the assembly. The silver solder is represented
  in this figure by grey pea-shaped objects next to the brass pucks. 
  The direction of helium flow is indicated by small black arrows 
  located inside the VCR glands. 
  }
\label{fig:cap_cross}
\end{figure}

The \spider capillary system achieves this using three 35~cm long
capillaries wrapped around Teflon spools and silver soldered into
1/8~inch \textit{Swagelok} glands with gender preserving VCR fittings
(see Figures \ref{fig:cap} and
\ref{fig:cap_cross}).\footnote{\textit{Swagelok}, Solon, OH.} Note
that Figure \ref{fig:cap} shows an assembly that is able to support
four capillaries; the fourth capillary was removed and replaced with
caps to reduce loading of the main tank. Having 3--4
capillaries reduces susceptibility to constrictions from potential ice
slush in the main tank. Failure of one capillary, most likely due to
an ice plug, should not reduce the performance of the superfluid tank
if safety factors are chosen appropriately.
The capillaries couple two $\sim$50~mL volumes that are connected to the main tank and the
superfluid tank through 1/8~inch bellows tubing. The capillary
material is extruded SS 304 with a 0.0035~inch inner diameter and a
0.0025~inch wall thickness.\footnote{Capillaries provided by \textit{Eagle 
Stainless}, Warminster, PA.} The dimensions were chosen to resemble
one of the assemblies described in \cite{Delong1970}. 

Small, custom-made, brass pucks with 0.009~inch diameter
center-drilled holes are made to fit snugly into the Swagelok
glands. A capillary is threaded through one of the pucks before solder
is allowed to flow into gaps that exist between the gland, puck,
and capillary. The capillaries are then wrapped around hollow Teflon
spools and subsequently routed back into the center of the spools and
through stainless steel compression springs that provide structural
support to the whole assembly while keeping heat conduction at a
minimum. As the capillaries are internal to the springs, there is
little chance that spring compression can pinch the capillaries. The
springs allow us to gently modify the overall length of the assembly
to fit perfectly between the two glands that are welded to each of the
small boxes while conducting negligible heat between the two
temperature stages. Transparent heat-shrink tubing helps align the
Teflon components with the two ends of the VCR fittings and provides
structural rigidity. Together, the capillaries along with VCR
fittings, Teflon components, and springs, are referred to as
capsules. Figure~\ref{fig:cap_cross} shows a cross section of an
individual capsule.

Stainless steel Mott filters are spot welded to the inside of the
MT~box so as to intercept large particles before they enter the
capillaries.\footnote{Filters purchased from \textit{Mott Corporation}, 
Farmington, CT. We use media grade 20 filters.}  
If correctly installed, the
particle capture efficiency of these filters is such that they collect
99.9\% of particles whose diameter is larger than 20\% of the
capillary diameter. This should allow for effective operations of the
capillaries without contributing significantly to the overall flow
impedance of the system.

Measurements show that the capillary assembly provides approximately
100~mW of cooling power to a superfluid volume while conducting only
2~mW between the two temperature stages. This corresponds to an
equilibrium flow through the capillaries of about $2.0\pm0.5$~SLPM.\footnote{
SLPM stands for Standard Liter Per Minute, where standard refers to 
273.15~K temperature and 1~atm pressure. For \hen, we find that 2.00 SLPM corresponds to 5.87~mg/s.} 
By boiling off the SFT and measuring the flow rate we estimate an
equilibrium superfluid liquid volume of approximately 2.0~L, only 1/8~of the
total volume of the SFT. This value is mainly determined by the
geometry and capillary flow rates.

The net cooling power can be changed by simply altering the length of
the capillaries in a way that does not require any other changes to
the design. With a collection of interchangeable capsules having
different flow impedances we can quickly arrive at an optimal cooling
power.

\subsection{Telescope Architecture and Adsorption Refrigerators}
\label{sec:telref}

\begin{figure}[tb]
\begin{center}
    \includegraphics*[width=0.45\textwidth]{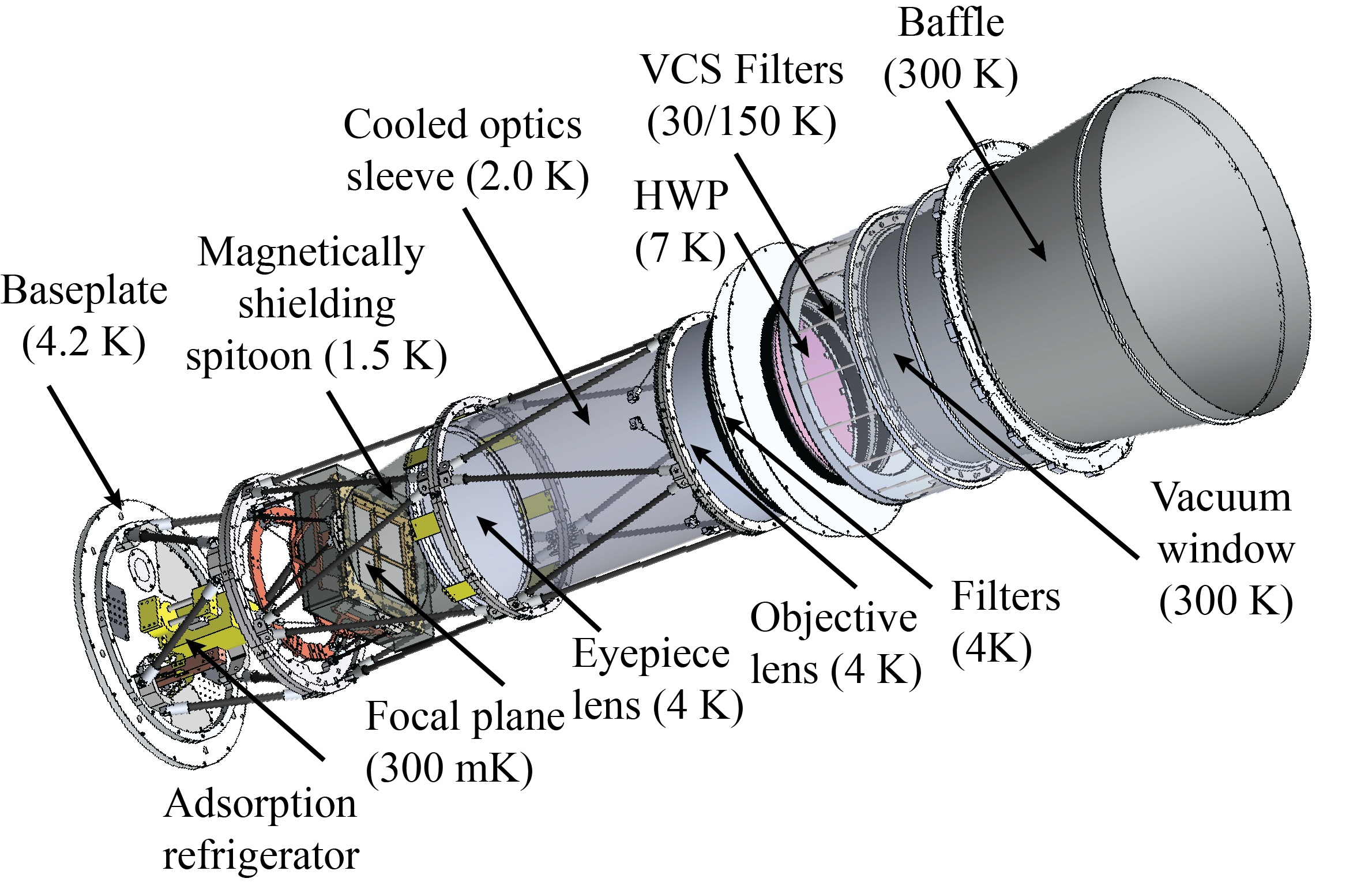}
    \caption[The \spider telescope assembly]{A CAD model of the entire
      telescope assembly, measuring 2.02~m end-to-end.  The objective lens,
      4~K filters, and half-wave plate are not clearly visible from
      this viewing angle. The VCS filters are omitted to reveal the
      surface of the HWP.}
\label{fig:telescope}
\end{center}
\end{figure}

The \spider telescopes are based on an optical design that is similar
to that of the \bicep telescope (see Figure
\ref{fig:telescope}) \cite{Yoon2006,Aikin2010}. Because of the small
aperture, the entire optics assembly can be cooled to 4~K, up to and
including the cryogenic waveplate mounted in front of the objective lens
(see Section \ref{sec:hwp}). A set of filters are mounted on both
vapor cooled shields (see Section \ref{sec:filters}) and a 1/8~inch
thick, mm-wavelength transparent, ultra-high-molecular-weight polyethylene window is mounted in front of the optics in a reentrant window bucket that is bolted to the top dome
of the vacuum vessel. The six telescopes that flew on \spidern's 2015
flight are referred to as X1--X6.

The telescope frame is constructed from carbon fiber structural
members epoxied into aluminum fixtures and mounted to four aluminum rings.
These four rings are connected to the baseplate, focal plane, 
eyepiece lens, and objective lens, respectively. 
The carbon fiber provides a conductive and
lightweight structure with outstanding rigidity. At the telescope 
operating temperatures (0.3--4~K), the carbon fiber 
supports have a relatively high ratio of elastic
modulus to thermal conductivity (approx. 10~MPa-K-m/W at 2~K) 
when compared to other polymeric and
composite structures, including G--10 \cite{Runyan2008}. Carbon fiber, 
however, can not be used for the structural flexures supporting
the main tank, as the thermal conductivity of carbon fiber becomes 
exceedingly large at temperatures above 4~K. For example, the ratio of
carbon fiber to G--10 thermal conductivity is approximately 2 and 300 at 20 
and 100~K, respectively.

A flexible copper shim heat
strap is routed internally to the telescope, creating a thermal link between the focal
plane and the still of the adsorption refrigerator that is mounted to
the baseplate. This heat strap ensures that the focal plane is cooled
to 300~mK.  Another heat strap connects the superfluid temperature
stage to the magnetically shielding spittoon and a cooled optics sleeve 
located between the two lenses. A third copper strap
provides a robust 4~K connection from the baseplate to the two lenses.
During flight, the temperature of the objective lenses was measured to
be 0.5--1.0~K above the temperature of the baseplates.

The telescopes use high magnetic permeability materials to reduce
susceptibility to magnetic disturbances. Informed by finite element
modeling, the focal plane, as well as the SQUID readout system are
surrounded by a 1.6~K spittoon made of \textit{Amumetal~4K} (A4K), a
proprietary nickel and iron alloy.\footnote{\textit{Amuneal
    Manufacturing Corp.,} Philadelphia, PA.} The SQUID mux chips are
further shielded by a combination of superconducting niobium backshort, 
a superconducting niobium
enclosure, and A4K sleeves \cite{Runyan2010,Filippini2010}. Finally,
the telescope tubes are surrounded by a two-layer concentric A4K or
\textit{Cryoperm~10} magnetic shield assembly which are expected to
equilibrate at approximately 5~K.\footnote{Cryoperm 10 was superseded
  by A4K during the fabrication process.}

Heat straps connect the superfluid tank to each of the telescope tubes,
providing a base temperature for the \linebreak 10~STPL closed-cycle $^3\mrm{He}$
adsorption refrigerators that cool each focal plane.\footnote{STPL
  stands for Standard Temperature and Pressure
  Liter.}${}^{,}$\footnote{Refrigerators supplied by \textit{Chase
    Research Cryogenics}, Sheffield, U.K.} Of order 10~g of activated
charcoal are installed inside the pump of a single refrigerator. When
cooled to 4~K, this charcoal is capable of adsorbing all gaseous
$^3\mrm{He}$ in the refrigerator. The adsorption refrigerator is
equipped with a helium gas-gap heat switch which can thermally link
the pump to the 4~K base temperature of the main tank. With the heat
switch open, and with the charcoal pump maintained at approx.\ 35~K, any
$^3\mrm{He}$ gas in the refrigerator will cool, condense, and
eventually fall into a small reservoir, located below
the condensation point, which is maintained near 1.6~K by the SFT. The
heat switch is closed 30--45~min after the charcoal pump has been
heated. This ensures that all 10 STPL are gaseous at the time when
liquefaction begins. As the pump cools, $^3\mrm{He}$ gas above the
liquid bath will begin to adsorb again, causing a reduction in
the vapor pressure of the $^3\mrm{He}$ bath and, therefore, the
temperature of the still \cite{Duband1990,Duband1991}. A successful
fridge cycle results in a 300~mK still temperature with hold times of
3--4 days, although we choose to cycle our fridges more frequently
than that (see Sections \ref{sec:heleak} and \ref{sec:cryobe}).

\begin{figure}[b]
\begin{center}
    \includegraphics*[width=0.49\textwidth]{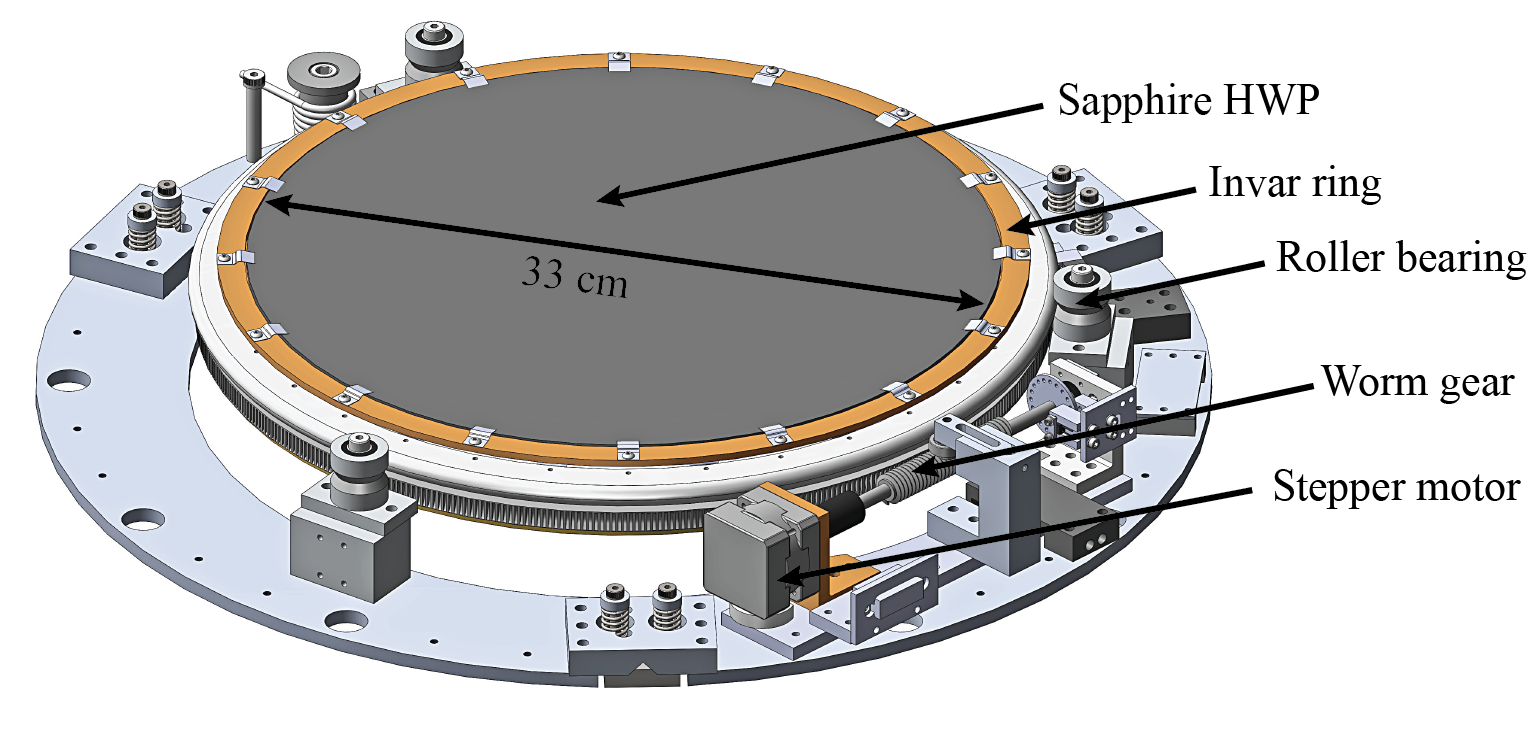}
    \caption[The \spider HWP]{A CAD model of the half-wave plate assembly for \spidern's 150~GHz telescopes. The worm gear and three bearings are also shown.}
\label{fig:hwp}
\end{center}
\end{figure}

\subsection{Cryogenic Half-Wave Plate}
\label{sec:hwp}

\spider uses birefringent sapphire half-wave plates \linebreak (HWP) to modulate polarization on the sky \cite{Bryan2010a}. This modulation, together  with sky rotation, differentiates a polarization signal from the sky from spurious instrumental polarization produced below the HWP. Each \spider half-wave plate is mounted on a gold-plated Invar ring with compression clips holding the sapphire in place (see Figure \ref{fig:hwp}).

A slightly customized cryogenic stepper motor (\textit{Mycom
  PS445-01A}) spins a worm-gear that couples to teeth on the plate
assembly with steady rotation facilitated by three roller bearings
that are distributed evenly around the plate.\footnote{Mycom
  Technology, www.mycom.com.sg} Quadrature optical encoders read out
the angular orientation of the sapphire plate. The half-wave plates
are mounted on top of the main tank, right in front of the objective lens
(see Figure \ref{fig:telescope}). Measurements suggest bearing and
motors equilibrate at around 7~K, but the weak thermal link between
the plate and the bearings, and therefore the main tank, causes the
sapphire plate and anti-reflection coatings to equilibrate closer to
30~K.

A single HWP assembly weighs 6.5~kg. During flight, each HWP was
rotated by a few tens of degrees every 12 sidereal hours. Measurements
suggest that stepping the six HWPs at this rate does not load the main
tank by more than 60~mW (1~L/day). This includes loading from
cryogenic wiring.

\subsection{Filters}
\label{sec:filters}

Radiative loading from the six 300~mm diameter telescope apertures
creates a difficult thermal shielding problem.
Millimeter-wavelength photons ($\nu < 300$~GHz) must pass through the
VCS apertures with good efficiency, while preventing higher-frequency
thermal radiation from the atmosphere, ambient-temperature window, and
warmer temperature stages from overloading the cooler cryogenic
stages.  We use a combination of reflective metal-mesh
\enquote{shaders} and low-pass filters \cite{Ade2006,Tucker2006} located at the
vacuum window, VCS2, and VCS1 to reduce the cryogenic loads to
acceptable levels.  Two shaders with a 1--2~THz low-pass edge are
suspended below the vacuum window to attenuate the majority (90\%)
of the infrared emission from both the atmosphere and the window itself.
Four such shaders at VCS2, followed by three more shaders and a
12~$\mrm{cm}^{-1}$ (360~GHz) low-pass filter at VCS1, further attenuate more than 95\% of the
incident radiative load at each stage.  Under float conditions,
we estimate about 9~W, 300~mW, and 3~mW of loading through all six
filter stacks to VCS2, VCS1 and the MT, respectively.
See further discussion of filter modeling in Section \ref{sec:thermal_model}.

\subsection{Cryogenic Wiring}
\label{sec:cryowire}

The \spider flight cryostat uses an ensemble of copper, Manganin, and
phosphor bronze wiring. Wiring for each telescope requires 400
conductors of 36~gauge Manganin wire, 300 of those are employed for
detector readout while the remaining 100 are reserved for general
telescope housekeeping, including thermometer and heater wiring. Each HWP
requires 32 conductors with a combination of 32- and 36-gauge phosphor
bronze wire. All conductors are heat sunk on VCS1, but not on VCS2, in
order to prevent a thermal short between the two shields. Copper is
only used internal to each temperature stage. 
Assuming the housekeeping cables are properly heat sunk at
VCS1, we expect a total of approximately 5~mW loading to the main tank through
wiring.

\section{Ground Operations and Characterization}

\begin{table}[]
\caption{Chronological list of evacuation and pre-cooling tasks. \label{tab:sched}}
\begin{center}
\begin{tabular}{ll}
\hline
\textbf{Task} & \textbf{Time}  \\
\hline
\begin{minipage}[t]{0.7\columnwidth} Multiple purge cycles  \end{minipage} & 36--48 hours \\
\begin{minipage}[t]{0.7\columnwidth} Pump with turbomolecular pump \end{minipage} & 3--4 days \\
\begin{minipage}[t]{0.7\columnwidth} Fill and equilibrate at $\mrm{LN}_2$ temperatures \end{minipage} & 4--5 days \\
\begin{minipage}[t]{0.7\columnwidth} Cool to LHe temperatures  \end{minipage} & 3 days \\
\begin{minipage}[t]{0.7\columnwidth}\textbf{Total} \end{minipage} & \textbf{11--14 days} \\
\hline

\end{tabular}
\end{center}
\end{table}

\subsection{Evacuation and Pre-Cooling}

\begin{figure}[tb]
\begin{center}
    \includegraphics*[width=0.49\textwidth]{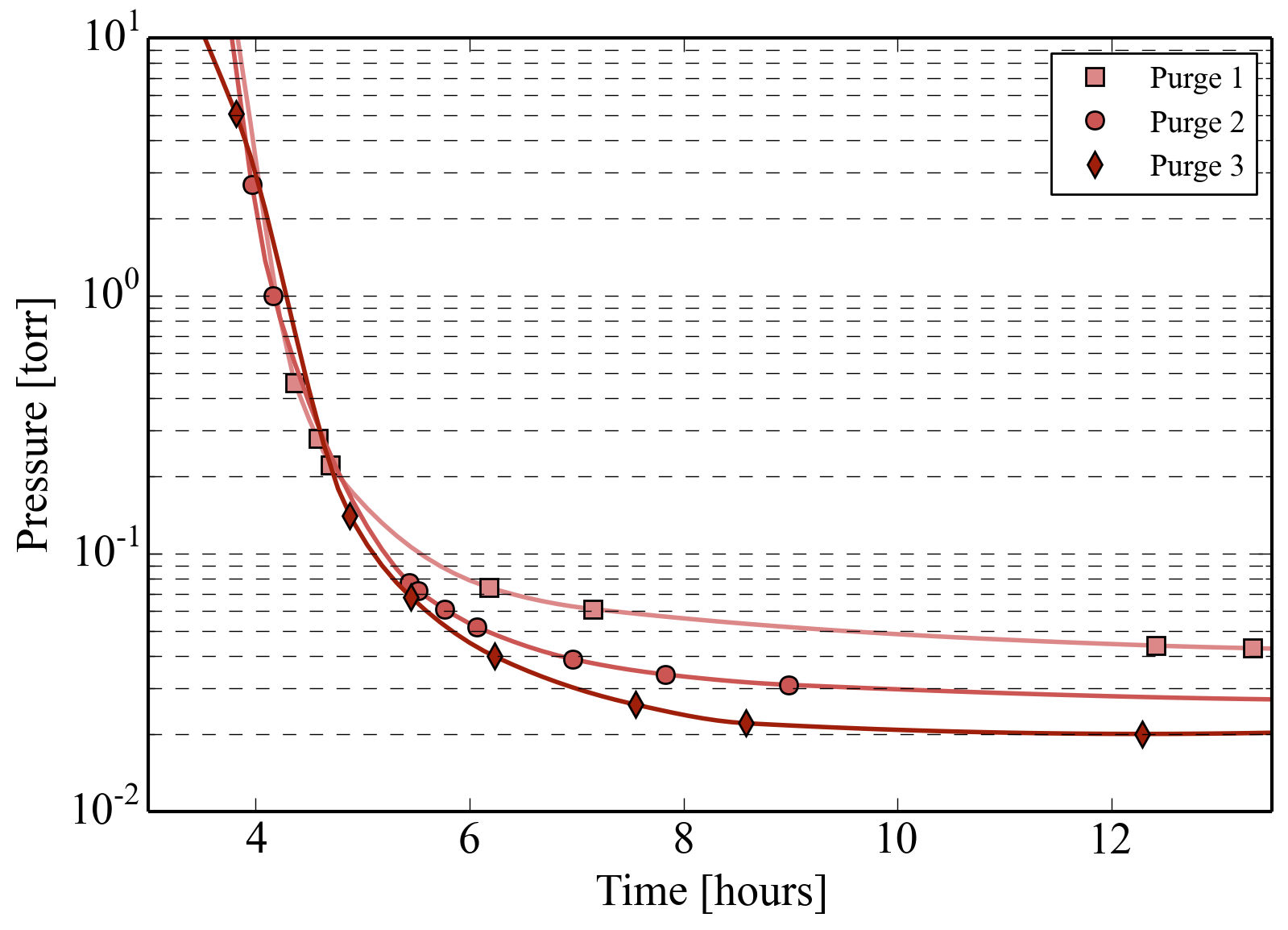}
    \caption[Pumpdown profiles]{Pumpdown profiles for the vacuum vessel pressure following three consecutive purge operations. Each purge further reduces the asymptotic pressure. Markers represent measurements, with lines added to guide the eye. The time-axis represents the time elapsed since the start of the pumpdown. These pumpdown profiles were observed in Antarctica before the pre-flight cooldown of the \spider cryostat. Prior to shipping, the cryostat had been purged with GN$_2$.}
\label{fig:press_pro}
\end{center}
\end{figure}

The vacuum vessel requires approximately one week of pumping before
the initial liquid nitrogen cooldown can begin. During the pumpdown
procedure, we find that backfilling with dry nitrogen helps to reduce
the asymptotic absolute pressure, presumably by freeing water vapor 
adsorbed on the interior surfaces. Our
experience suggests three to four gaseous nitrogen (GN$_2$)
backfilling procedures are useful, but any subsequent purge does not
seem to help reduce the asymptotic pressure. We find that purging to 
higher pressure (760~torr rather than 100~torr) is generally more 
effective, with the caveat that time is also a significant parameter in this process. 
We use a dry scroll pump (Agilent TriScroll 600)
for this initial \enquote{rough-out} phase of the
pumpdown.\footnote{\textit{Agilent Technologies}, Santa Clara, CA.}
Figure \ref{fig:press_pro} shows the pressure profiles following three
consecutive purge operations with the pumpdown manifold fixed 
between purges. The reduction in asymptotic 
pressure values suggests the purge operations are helpful.

After this initial phase we install a turbomolecular \linebreak pump directly at
the KF-50 pump-out flange on the vacuum vessel so as to minimize the
flow impedance between the vacuum vessel and the pump. When moving the
pump further away from the pumpout flange, we observe significant
degradation in pumping speed in accordance with the expected inverse length scaling.

When the pressure reading at the vacuum vessel \linebreak pumpout
port indicates approximately $2\times10^{-4}$~torr (while pumping),
we can begin the liquid nitrogen pre-cooling. This process removes
90\% of the combined enthalpy of the telescopes and cryostat, requiring
300--400~L of liquid nitrogen. The subsequent cooldown to 4~K consumes
approximately 500~L of liquid helium before the cryostat has fully
equilibrated. The cooldown operation is performed by filling with
100--200~L of liquid a few times over a 48 hour time
period. Table~\ref{tab:sched} highlights the main aspects of the
evacuation and cooldown schedule and their approximate duration. The
considerable time requirement is set by the net volume of the vacuum
vessel, approximately $6\:\mrm{m}^3$, as well as the total mass that
is cooled to the 4~K base temperature, about 240~kg from the main tank
and an additional 48~kg per telescope (an approximate total of 550~kg is cooled to 4K).

During liquid nitrogen pre-cooling and beginning period of helium
cooling, the thermally isolated capillary assembly is maintained at
300~K using heaters mounted to the SFT box. The SFT is also
pressurized to 21~psia with high-purity helium gas in order to
prevent flow of liquid nitrogen into the capillaries, potentially
causing flow restrictions. After a significant amount of liquid helium
has accumulated in the main tank and the VCSs are nearing their
equilibrium temperature, we turn off heat to the capillary assembly
and begin pumping on the SFT vent line. This initiates the flow of
liquid helium through the capillaries, becoming superfluid in the
process, with its evaporation providing cooling power to the SFT. With
the capillaries heated to 300~K during LN$_{2}$ operations, we have
not yet observed a flow restriction in the capillaries when cooling
the SFT down to operational temperatures. The SFT reaches superfluid
temperatures about 18 hours after turning off heat to the
capillaries. We can typically cycle fridges within 24 hours of the SFT
temperature falling below 2~K.

\subsection{Helium Leak}
\label{sec:heleak}

The main and superfluid tanks of the \spider flight cryostat each have
at least one microscopic fissure that leaks helium. The leaks were
first detected during the construction phase and we have had mixed
success in locating and sealing them. The presence of the leaks can be
detected at room temperature, but only by pressurizing the tanks
with gaseous helium while pumping on the vacuum vessel; the leak rates
are minuscule, making then impossible to localize at room
temperature. The leaks become more apparent as the cryostat is cooled
to liquid nitrogen temperatures, and early attempts at localizing them
by covering the affected area with liquid nitrogen proved successful.
The involved network of flow impedances between the leak and the vacuum vessel 
pumpout port prevents accurate determination of the absolute
leak rates.

We have applied \textit{Stycast 2850FT} to suspect areas and noticed a
marked reduction in helium backgrounds, but also found that these epoxy
seals can worsen with time. The large number of aluminum welds (17~m
in total) as well as the complex shape and high density of threaded
holes make this design relatively susceptible to manufacturing errors
resulting in micro-fissures. Our results suggest that the aluminum
welds are the most likely culprits. Based on experience dealing
with recurring leaks, we decided to apply Stycast to all suspect areas
during the pre-flight closeup in November 2014, with particular
attention to the welds.

The increased helium background in the vacuum vessel can negatively
affect the performance of the transition edge sensors populating each
focal plane, in addition to adding a parasitic heat load to the cryogenic
stages. The prevailing theory posits that a superfluid film
builds up on the focal planes, effectively changing the thermal
properties of the detectors and therefore both their gain and the
noise properties. This phenomenon is time-dependent, has an abrupt onset, 
and a predictable cycle. In order to
negate this build-up of superfluid film we cycle the fridges more
often than required based on the hold-time alone (see Sections
\ref{sec:telref} and \ref{sec:cryobe}) as the fridge cycles clear the films. 
The result is a slight decrease in duty cycle relative to the design, 
with no other impact on detector performance.

\subsection{Results from Ground-Testing}
\label{sec:gebe}

The Antarctic long-duration balloon program has launc\hyp{}hed 
approximately 40 payloads to date, and
the average flight time is 20~days \cite{Gregory2004,csbf_traj}. 
Based on this observation, we defined a 20~day hold time
requirement during the design phase of the \spider flight cryostat.

A lab measurement with all six telescopes mounted, flight-ready, 
and observing a room temperature load, shows a 40~SLPM
average flow from the main tank.  
This corresponds to approximately $82~\mrm{L}$ of daily
liquid loss and 12 days of cryogenic hold time assuming an initial
$1,000~\mrm{L}$ charge of liquid helium. During these tests, the two
vapor cooled shields, VCS1 and VCS2, equilibrate at 42 and 157~K,
respectively, with approximately 2~SLPM of helium gas flow from the
main tank to the superfluid tank. Similar cryogenic tests that were
conducted with both vapor cooled shields fully sealed (filters
replaced with aluminum plates and covered in MLI), and without the
science instruments installed, showed significantly lower flow rates
and loading. During those tests we observed an average main tank flow
rate of 23~SLPM with VCS1 and VCS2 equilibrating at 28 and 157~K,
respectively. We attribute the difference in loading primarily to
optical power penetrating the telescope filter stacks. Table \ref{tab:cryoperf}
summarizes observations from cryogenic testing.

\begin{figure}[tb]
\begin{center}
    \includegraphics*[width=0.49\textwidth]{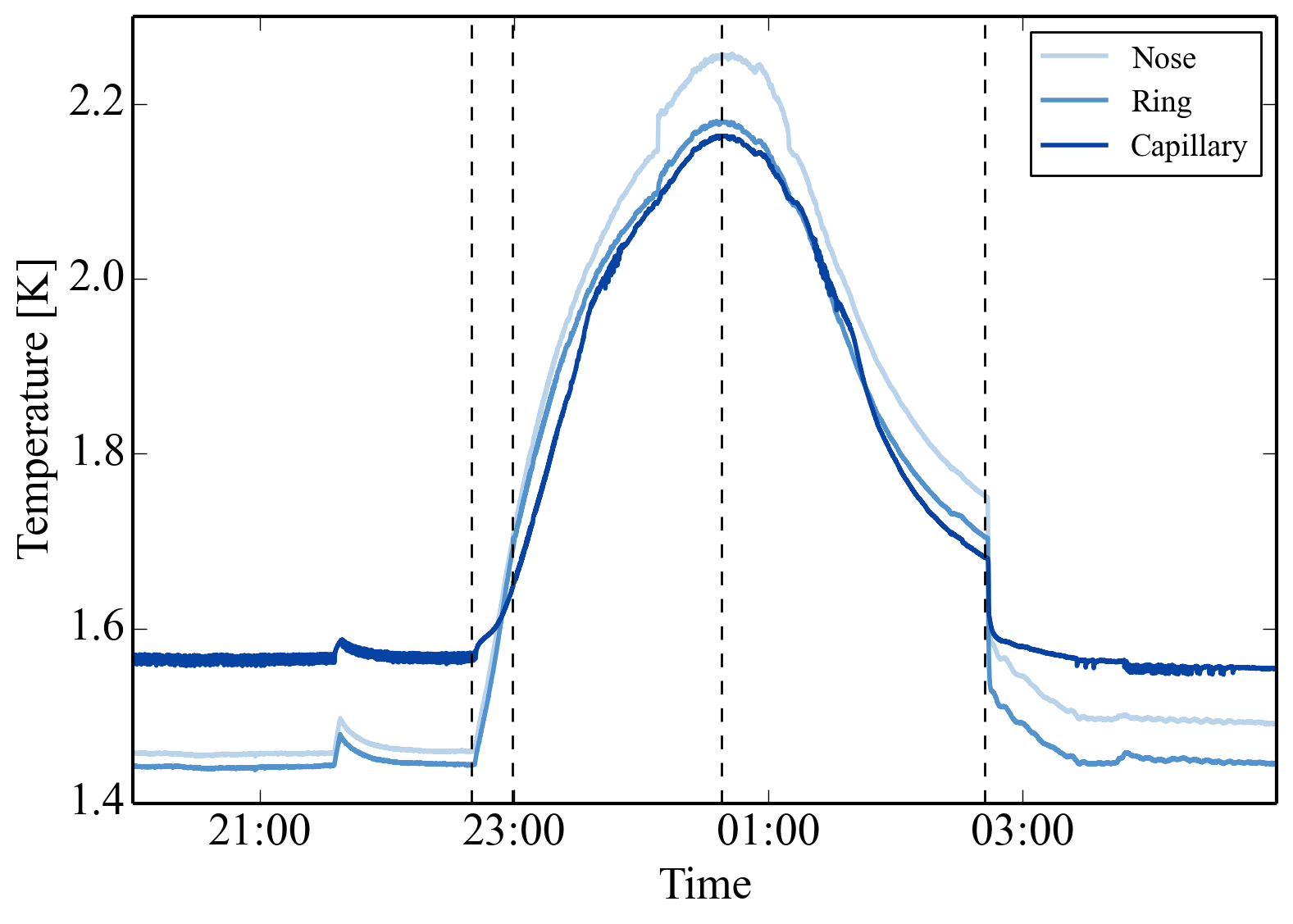}
    \caption[SFT launch transients]{The SFT temperatures during a
      9~hour period that includes \spidern's launch. For this, and all
      other plots showing time development, we use Eastern Time
      (ET). In this time zone, the payload was launched just an hour
      before midnight on New Years Eve. From left to right, the four
      vertical dashed lines represent the time when: 1) the Triscroll
      pump that is pumping on the SFT is replaced with a smaller
      diaphragm pump; 2) \spider is launched; 3) The SFT reaches its
      maximum temperature of about 2.2~K while the payload is at an
      altitude of 17~km where the ambient pressure is approximately
      59~torr; and 4) A motorized valve is opened, letting the atmosphere
      pump on the SFT and resulting in an immediate drop in
      temperature.}
\label{fig:sft_launch}
\end{center}
\end{figure}

\begin{table}[]
\caption{A list of equilibrium behavior during distinct cryogenic
  tests. VCS1 and VCS2 represent the equilibrium temperature of the
  two intermediate stages while flow stands for the helium flow out of
  the VCS vent line. Typically, about 3 and 15~K temperature gradients
  exist across VCS1 and VCS2, respectively. Columns labeled LN$_2$
  and LHe$^{\mrm{dark}}$ represent performance observed with all
  apertures on the vacuum vessel and the VCSs closed with aluminum
  plates, significantly less cryogenic wiring, and no telescopes
  installed. Conversely, LHe$^{\mrm{light}}$ and Flight represent
  performance observed with all six telescopes and flight-wiring
  installed. We note that 20.0~SLPM corresponds to 1.23~W of heat to
  the main tank and a mass flow of 58.7~mg/s.}
\label{tab:cryoperf}
\begin{center}
\begin{tabular}{lllll}
\hline
 & \textbf{LN$_2$} & \textbf{LHe$^{\mrm{dark}}$} & \textbf{LHe$^{\mrm{light}}$} & \textbf{Flight} \\
\hline
VCS1 [K] & 120 & 28 & 42 & 30 \\
VCS2 [K] & 220 & 157 & 157 & 118 \\
Flow [SLPM] & 2.5 & 23 & 38 & 28 \\
\hline
\end{tabular}
\end{center}
\end{table}

\subsection{Cryostat Thermal Modeling}
\label{sec:thermal_model}

A thermal model was developed to inform the equilibrium loading and
performance of the \spider flight cryostat both in flight and during
ground tests in various configurations. The model accounts for
conductive loads via the support flexures, wiring, and plumbing, gas conduction
within plumbing lines, the feedback from the enthalpy of the helium
vapor and several sources of radiative loading. Conductive and radiative loading
between the bulk vapor-cooled shields is modeled using the Keller
model \cite{KellerMLI} for multilayer insulation.

Radiative loading along the optical path of each telescope is modeled
using measured reflection, transmission, and bulk absorption
properties of filter elements. The total power incident on any
temperature stage is the sum of 1) incident radiation from the CMB and
ambient atmosphere and 2) emission from each of the skyward filters,
all attenuated by the net transmission through the intervening
filters. 

\begin{figure}[tb!]
\begin{center}
    \includegraphics*[width=0.49\textwidth]{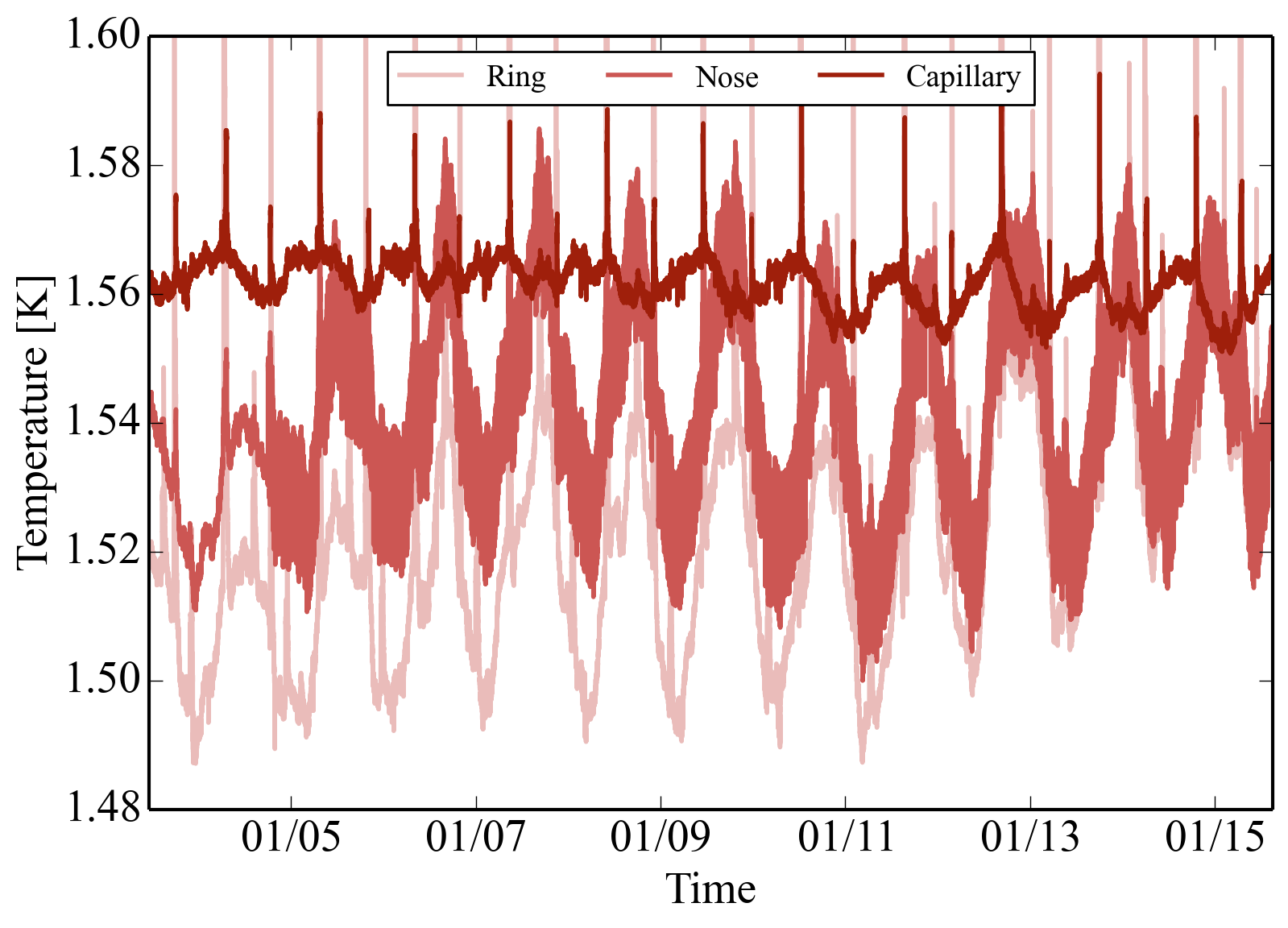}
    \caption[The SFT temperature during flight]{The temperature of the
      superfluid tank during the majority of flight. The ring and nose
      thermometers are located on the middle of the ring and the lower
      part of the bulk of the tank (see Figure \ref{fig:sft_cap}). The
      capillary thermometer was mounted to the SFT box. Spikes in
      temperatures are due to fridge cycles and diurnal variations are due to altitude, and therefore pressure fluctuations (approximately 40~$\mrm{\mu K/m}$).}
\label{fig:sft}
\end{center}
\end{figure}

Negative feedback is accounted for in the cooling power provided by
enthalpy of the boil-off gas through the heat exchangers to the
intermediate stages. Equilibrium temperatures and loads are found by
iteratively solving for zero heat flow across the vapor-cooled stages.
During flight-like conditions, VCS1 and VCS2 are expected to
intercept approximately 5 and 20~W, respectively.

In practice, our thermal model struggles to reproduce the observed main
tank loading under all operational conditions \cite{Gudmundsson2010}.
Careful accounting of conductive and radiative loading to
the main tank suggests upper limits of 800~mW and 200~mW,
respectively, when telescope apertures are replaced with aluminum
covers (corresponding to column LHe$^{\mrm{dark}}$ in Table
\ref{tab:cryoperf}). In this scenario, about 400~mW of power input to
the helium bath is missing from the model. A macroscopic bolometer
instrumented with an ensemble of heaters and thermometers was mounted
on the main tank to help estimate the radiative environment. Using
this bolometer we calculated a 200~mW upper limit to the radiative
loading on the main tank, which is consistent with the thermal model
predictions. However, the bolometer is not sensitive to light-leaks
that would most likely penetrate openings in MLI surrounding apertures
and main flexures. We suspect that unmodeled light-leaks might help to
explain the discrepancy between model predictions and realized
performance and note that incorporating such a feature into the
thermal model does improve consistency with observations. We also
note that interstitial residual helium, due to a leak,  
can increase the effective conductivity 
of MLI in ways that are difficult to model (see Section \ref{sec:heleak}). 

Pressure oscillations in plumbing lines can in some cases add significant 
thermal loads to cryogenic systems \cite{Taconis1949}. Careful heat sinking of the MT and SFT
fill and vent lines hopefully ensure smooth thermal gradients from 300 to 4~K (see Section \ref{sec:flexnplumb}). In-flight pressure data from the valved-off MT vent line, shows that the amplitude of  0.1--30~Hz pressure oscillations does not exceed 0.005~torr/rtHz with the exception of a narrow line at 0.04~torr/rtHz at 3.2~Hz.

Despite the excess of the observed loading above the model predictions, 
it is worth noting that \spidern's cryogenic performance is 
nonetheless impressive.  A na\"{i}ve scaling of effective areas from 
the test cryostats (including \biceptwon) to the \spider flight cryostat 
would predict a load double that which \spider actually experiences.

Despite significant reduction in atmospheric loading, a stratospheric payload is still subject to Earth's atmosphere. The pressure and temperature at float altitude is approximately 5 torr and 250 K, respectively. These conditions require a vacuum vessel, which inevitably intercepts of order 100 W of radiative loading per square meter. In space, passive cooling can reduce the base temperature of payloads down to $\sim$50~K \citep{planck2011-1.1}. This significantly reduces loading to the first active cryogenic state relative to a stratospheric payload.

The mass of the entire cryogenic assembly, including cryogens, 
was 990~kg. With a total cryogen volume of 1300~L, the 
effective density of the cryostat is 0.76~$\mrm{g/cm}^3$.  
Using a metric that combined depletion rate, $R$, with radiative 
loading, $H$, we find that $H/R \approx$ 60~W~day/L \cite{Holmes2001}.
This compares favorably to cryostats designed for space.

\section{Flight Performance}

\subsection{Pre-flight Configuration, Launch, and Ascent}
\label{sec:launch}

A number of cryogenic operations both shortly before launch and during ascent
are critical for a successful flight. \spider rolled out to the launch pad on January 1, 2015, and
successfully launched about nine hours later. \spider was launched
cold, with the detectors at 300~mK and the superfluid tank maintained
at low pressure with an auxiliary pump mounted on the launch vehicle.
At launch, the entire payload weighed 3065~kg (including everything from 
the pivot and below).

Ground operations require a dedicated pump to maintain the superfluid
tank below 2~K. At float, the SFT is open to the atmosphere, which at
36~km altitude easily reduces the vapor pressure of helium below the
$\mrm{\lambda}$-point. During the 3--4 hours required for ascent, we
pumped on the superfluid volume with a low-power \textit{KNF
  N950.50} diaphragm pump driven by one of our two 24~V flight battery
systems.\footnote{\textit{KNF Neuberger Inc.}, Trenton, NJ.} The pump
drew approximately 1~A at 24~V, providing a 2~SLPM pumping rate at
50~torr and exceeded our expectations from ground testing, presumably
because of reduced backing pressure at the pump outlet. Ground based 
testing suggested this pump would prevent the thermal
transients associated with valving off the superfluid tank entirely
while letting it pressurize due to the constant flow of helium from
the MT to the SFT.

\begin{figure}[tb]
\begin{center}
    \includegraphics*[width=0.49\textwidth]{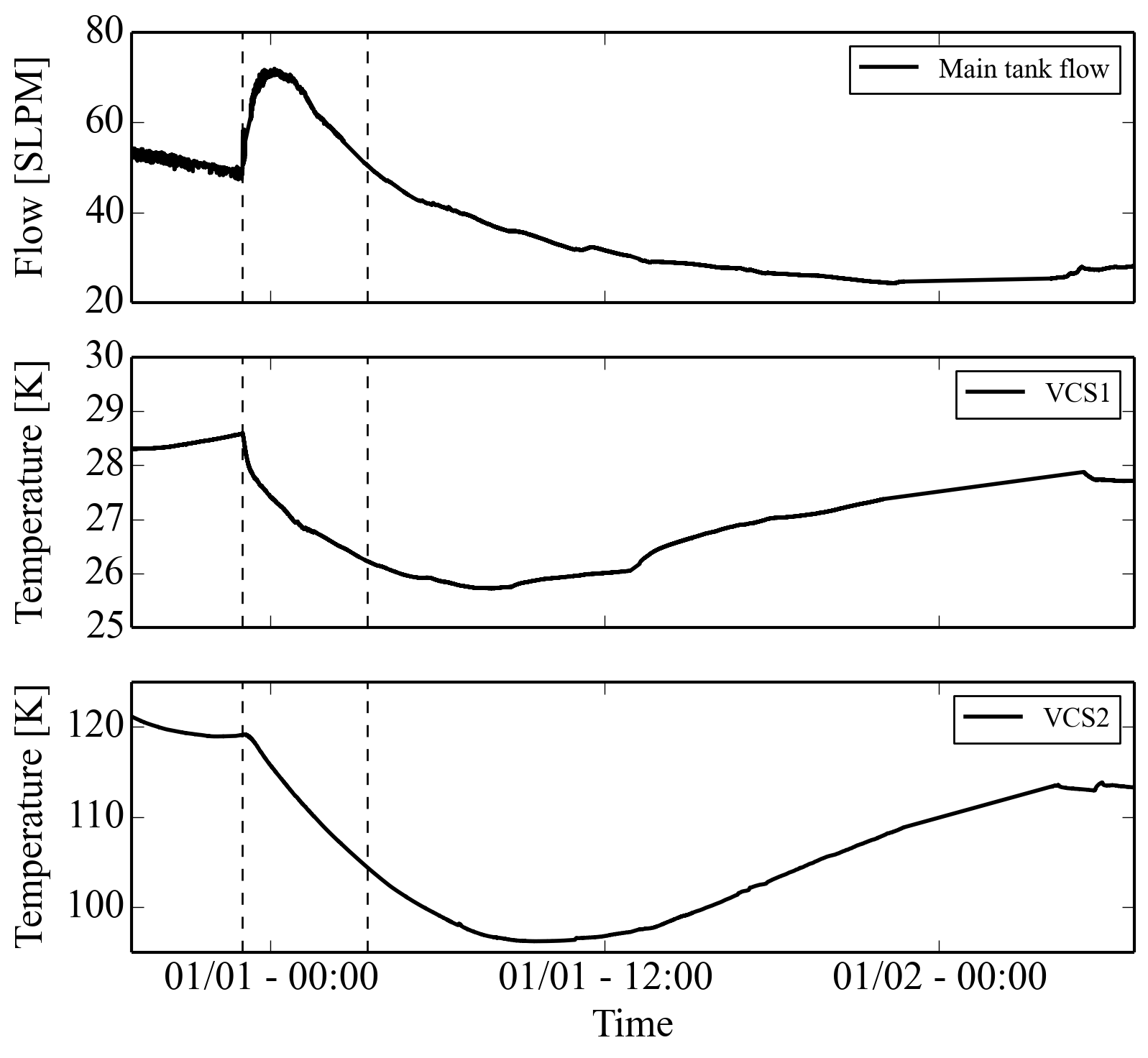}
    \caption[The MT flow and VCS temps]{The main tank flow and VCS
      temperatures during a 36~hour period that that includes
      launch. The significant rise in main tank flow coincides with
      the launch of the payload. The launch is followed by a 28~hour
      transient in the temperature of the vapor cooled shields. By the
      end of this time period, the main tank flow rate is
      equilibrating to its 28~SLPM nominal flow rate. Note that the
      cryostat had not fully equilibrated from the final helium fill
      when it was launched. The two vertical dashed lines indicate 
      launch and float altitude equilibrium at 36.4~km, respectively.}
\label{fig:flownvcs}
\end{center}
\end{figure}

Approximately 10~min before launch, we commanded a
motorized bellows valve shut, isolating the SFT vent line from a scroll
pump located on the launch vehicle. The diaphragm pump was then turned
on and allowed to pump on the superfluid volume during launch and
ascent. During this time, the temperature of the SFT only barely
exceeded the $\mrm{\lambda}$-point. Figure \ref{fig:sft_launch} shows
the temperature of the superfluid tank during a nine hour period that
includes \spidern's launch.

About 3.7~hours into launch, and with the payload 33.3~km above sea
level, we commanded a motorized valve to open the SFT vent line to the
atmosphere and subsequently turned off the diaphragm pump. Figure
\ref{fig:sft} shows the temperature profile of the SFT and capillaries
during flight. The tank was stable at 1.5--1.6~K for the entire
flight.  \spider equilibrated at a float altitude of $\sim$36.4~km,
approximately 4.5~hours after launch; the majority of the flight
was spent at an altitude of $35.5\pm0.5$~km, with the variance
mostly driven by diurnal fluctuations. At this altitude our ambient
pressure sensors recorded $4.7\pm0.3$~torr.

\begin{figure}[tb]
\begin{center}
    \includegraphics*[width=0.49\textwidth]{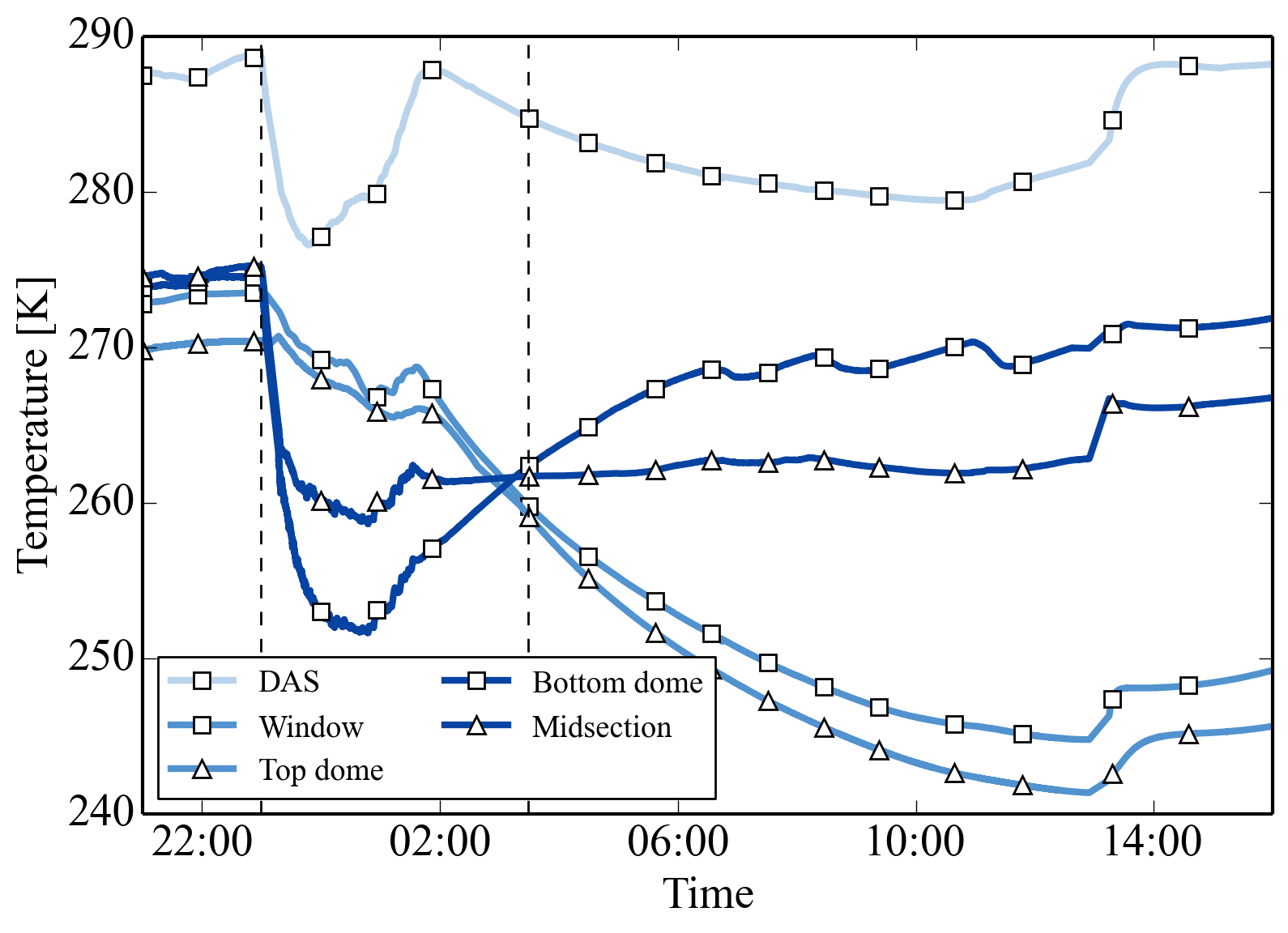}
    \caption[Vacuum vessel temperature during ascent]{The temperature
      of the vacuum vessel during ascent. The initial cooling
      transient results from the strong atmospheric temperature
      inversion in the troposphere.  The vacuum window and top dome
      cool radiatively to space while the bottom dome and midsection
      are almost surrounded by the sun-shields. The data-acquisition
      system (DAS), mounted to the midsection of the vacuum vessel
      just above the hermetic feedthroughs, cooled significantly on
      ascent, but recovered quickly. The two vertical dashed lines indicate 
      launch and float altitude equilibrium at 36.4~km, respectively.}
\label{fig:vv_ascent}
\end{center}
\end{figure}

Two Tavco absolute pressure regulators set to vent at 13.5~psia were
installed in parallel at the VCS vent where all gas exiting the MT is
routed. After exiting the vacuum vessel, but before passing through
these regulators, the cold helium gas passes through a heat exchanger
block which is thermally connected to the power supplies for the
multichannel electronics used to read out our detector signals
\cite{Dobbs2009}. This heat exchanger, a critical source of cooling
for the power supplies, which operate in near vacuum, provides approximately 
40~W of power to the cold gas. We employ two Tavcos at the VCS vent line to
split the flow and therefore reduce the effective cooling power and
the physical stresses experienced by each regulator. Given the
system's flow impedance and the estimated equilibrium flow, we
expected that the main tank would equilibrate at approximately
14.7~psia (1~atm) at float. Our in-flight pressure gauges suggest that
the main tank pressure equilibrated at $14.5\pm0.1$~psia. Another
Tavco regulator set to crack at 17.5~psia was installed on the MT vent
line in case of a sudden over-pressurization. That regulator only
opened at the conclusion of the flight, when the sudden loss of 
cooling power from evaporation of liquid helium resulted in 
transient heating of residual vapor.

During ascent, the atmospheric pressure dropped
rap\hyp{}idly. This resulted in a significant change in the pressure
differential driving flow out of the main tank through the vapor
cooled shields. This abrupt change in the pressure boundary conditions
caused a temporary increase in flow out of the VCS, which in turn
super-cooled the shields. 
Figure \ref{fig:flownvcs} shows the flow and VCS temperatures
during a 36~hour period that includes the launch. Note that 40.0 SLPM
corresponds to 2.46~W of loading to the main tank.

\begin{figure}[tb]
\begin{center}
    \includegraphics*[width=0.49\textwidth]{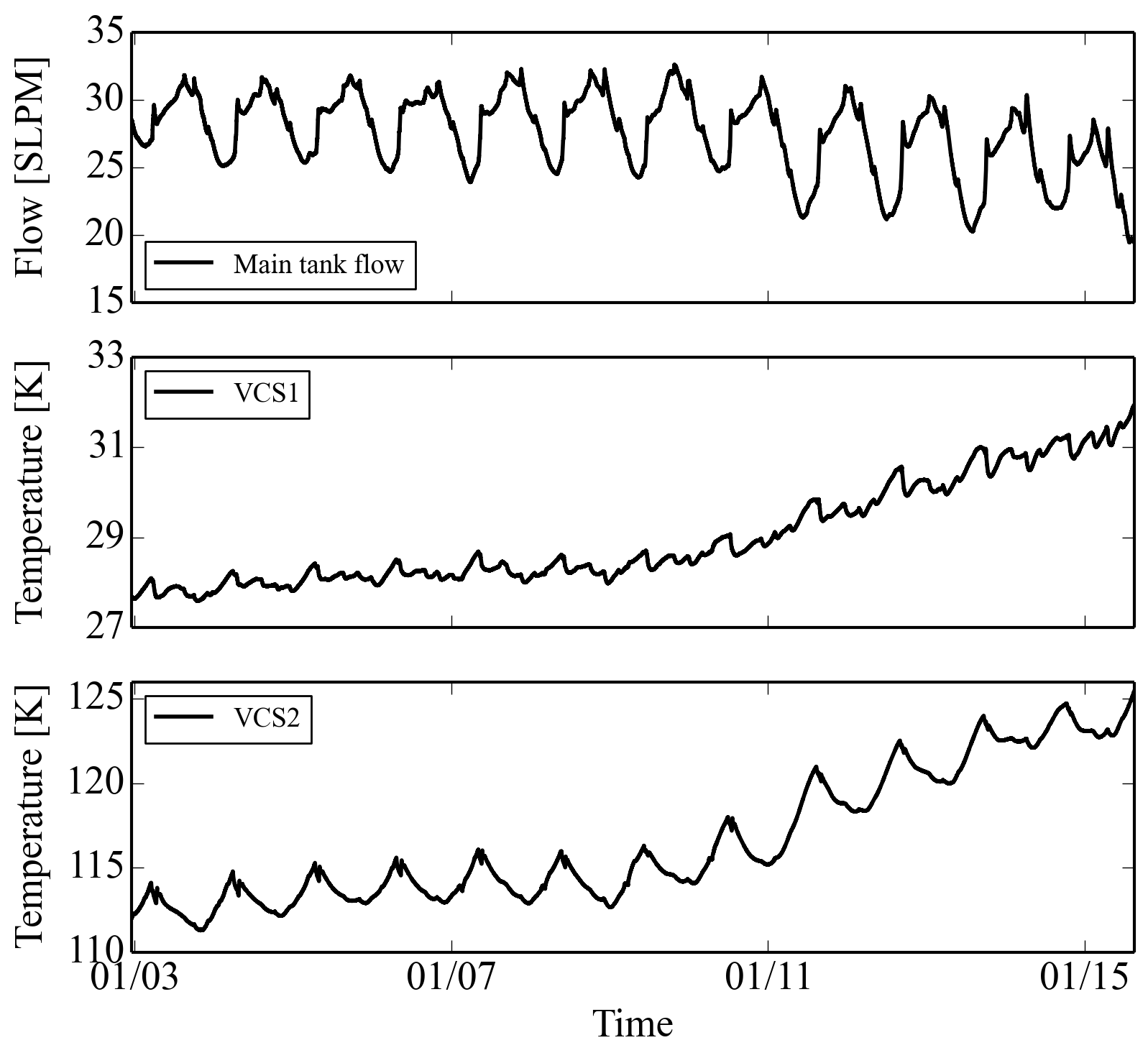}
    \caption[The MT flow and VCS temperatures]{The main tank flow and VCS
      temperatures over a majority of the flight duration. The steady decrease/increase in flow/VCS temperatures shows how the equilibrium loading depends on the main tank liquid level. Large diurnal variations from a combination of altitude and VV shell temperature variations are apparent.}
\label{fig:flownvcs_long}
\end{center}
\end{figure}

The payload surfaces experienced a significant temperature change as
the balloon ascended through the troposphere. Figure
\ref{fig:vv_ascent} shows the temperature at a few key locations of
the vacuum vessel. The largest transient lasted for about two~hours
during ascent and can be most clearly seen by the temperature of the
bottom dome and midsection of the vacuum vessel. About 15~hours after
launch, the vacuum vessel began to equilibrate to its nominal temperature.

\subsection{In-Flight Thermal Behavior}
\label{sec:cryobe}

\begin{figure}[tb]
\begin{center}
    \includegraphics*[width=0.49\textwidth]{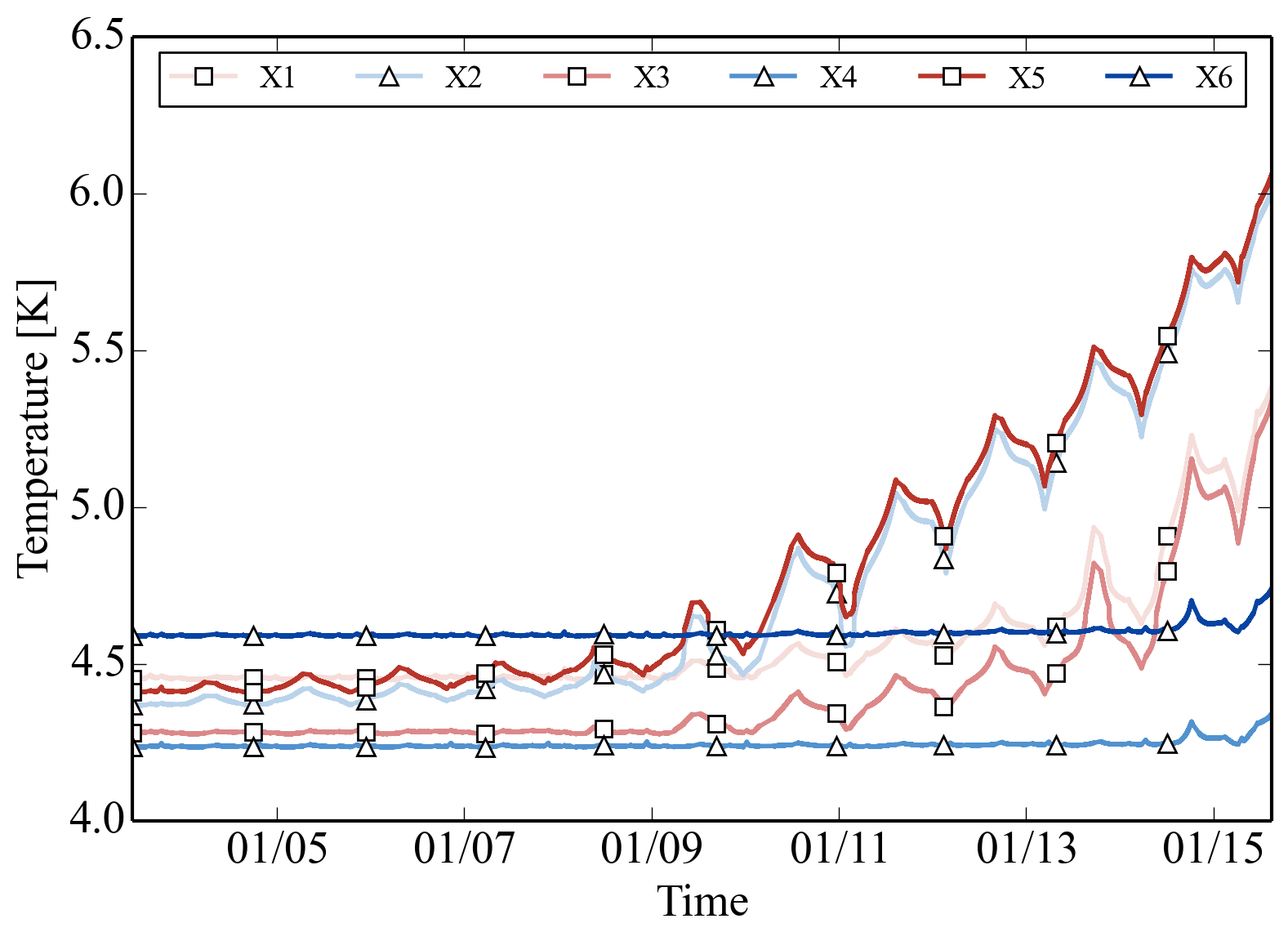}
    \caption[The baseplate temperatures]{The temperature profiles of
      the six telescope baseplates over the majority of flight
      duration. Brief temperature excursions caused by fridge cycles
      are omitted from the timelines for improved clarity.}
\label{fig:plates}
\end{center}
\end{figure}

Approximately 28 hours after launch, the flow out of the main tank had
stabilized to 28~SLPM, corresponding to an equilibrium loading on the
MT of about 1.7~W and a 58~L/day loss. This represents a 0.8~W reduction in loading
compared to that seen on the ground and is caused by a decrease in
vacuum vessel temperature as well as less sky loading. Table \ref{tab:cryoperf} compares
the in-flight equilibrium flow and VCS temperatures with that measured on the ground.
Diurnal and fridge cycling variation resulted in a 3~SLPM root mean square
variations in the flow rate on a daily basis. We note that the loading
at float is still greater than what we see on the ground with all
apertures blanked off (see Section \ref{sec:gebe}), suggesting that
radiative loading through telescopes contributes more significantly to
the MT loading than a 25~K reduction in the temperature of the vacuum
vessel shell. Figure \ref{fig:flownvcs_long} shows the steady state main tank flow 
and VCS temperatures during the majority of flight.

\begin{figure}[t]
\begin{center}
    \includegraphics*[width=0.49\textwidth]{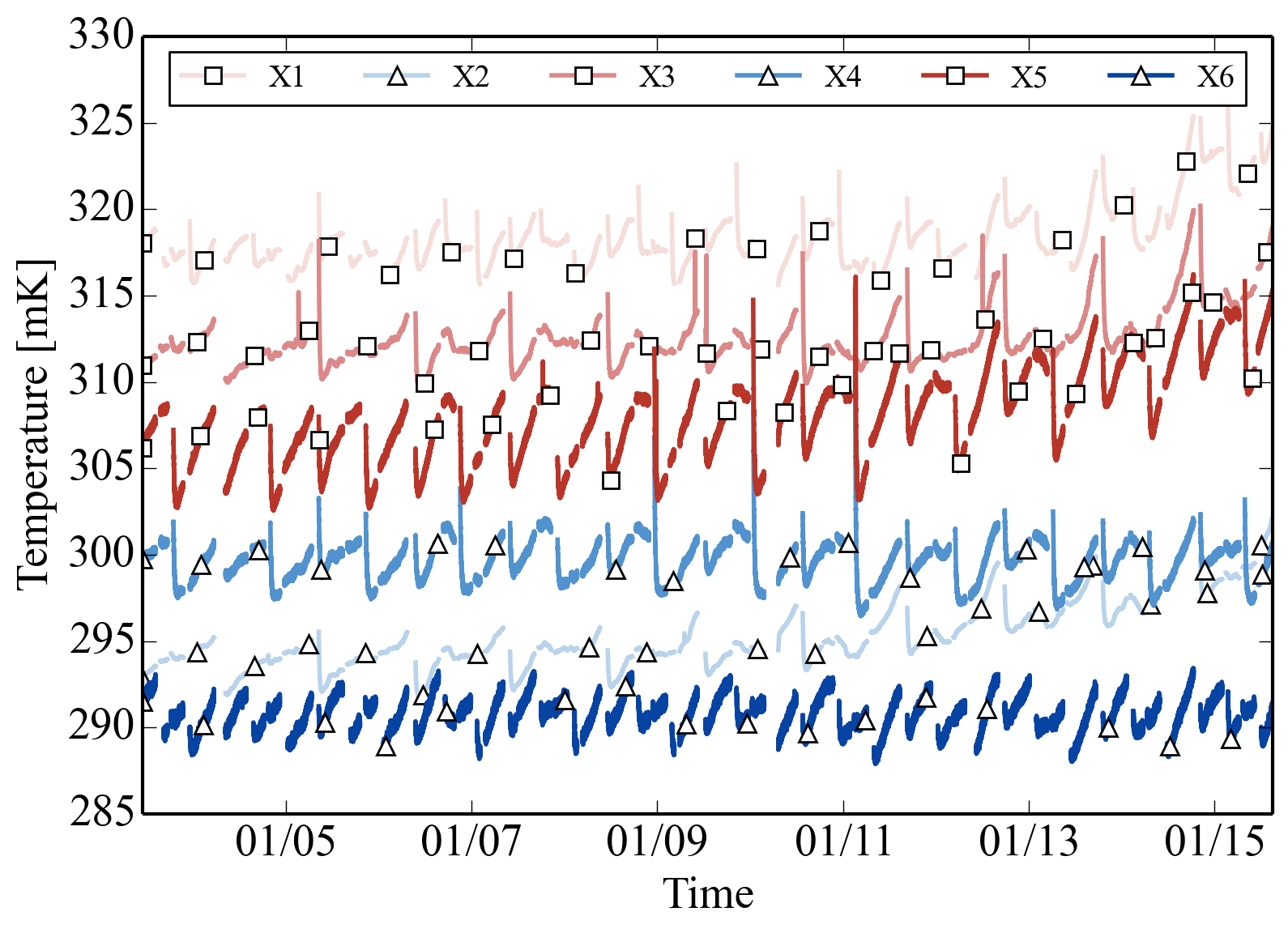}
    \caption[The FPU temperature]{The temperature variation of the six
      focal planes for the majority of the flight. Large temperature
      excursions due to fridge cycles are omitted and some 
      focal plane temperatures have
      been translated by approximately 10~mK to improve clarity. Note
      that four of the six focal planes show a slow drift in
      temperature as the cryostat comes close to depletion.}
\label{fig:fpu}
\end{center}
\end{figure}

The six telescopes operated nominally for the duration of the
flight. As expected, the baseplates warmed up with time as liquid
level decreased, but their temperatures remained suitable for
operations throughout the flight. Figure \ref{fig:plates} shows the
temperature profiles of the six telescopes baseplates over the
majority of the flight duration. Telescopes X2 and X5 were mounted in
the ports furthest from the bottom of the main tank (see Figure
\ref{fig:mtsft}), and therefore rose most significantly in temperature
with decreasing liquid level. Telescopes X1 and X3 were mounted in the
two middle ports and also saw some temperature rise as liquid level
dropped while the telescopes mounted in the bottom two inserts, X4 and
X6, did not change significantly in temperature until the cryostat
began to run out of liquid, around January 16. Strong thermal
connections between the baseplate and telescope ensured that
components such as the eyepiece and objective lenses did not deviate
from the baseplate temperatures by more than 0.5--1.0~K.

\begin{figure}[t]
\begin{center}
    \includegraphics*[width=0.49\textwidth]{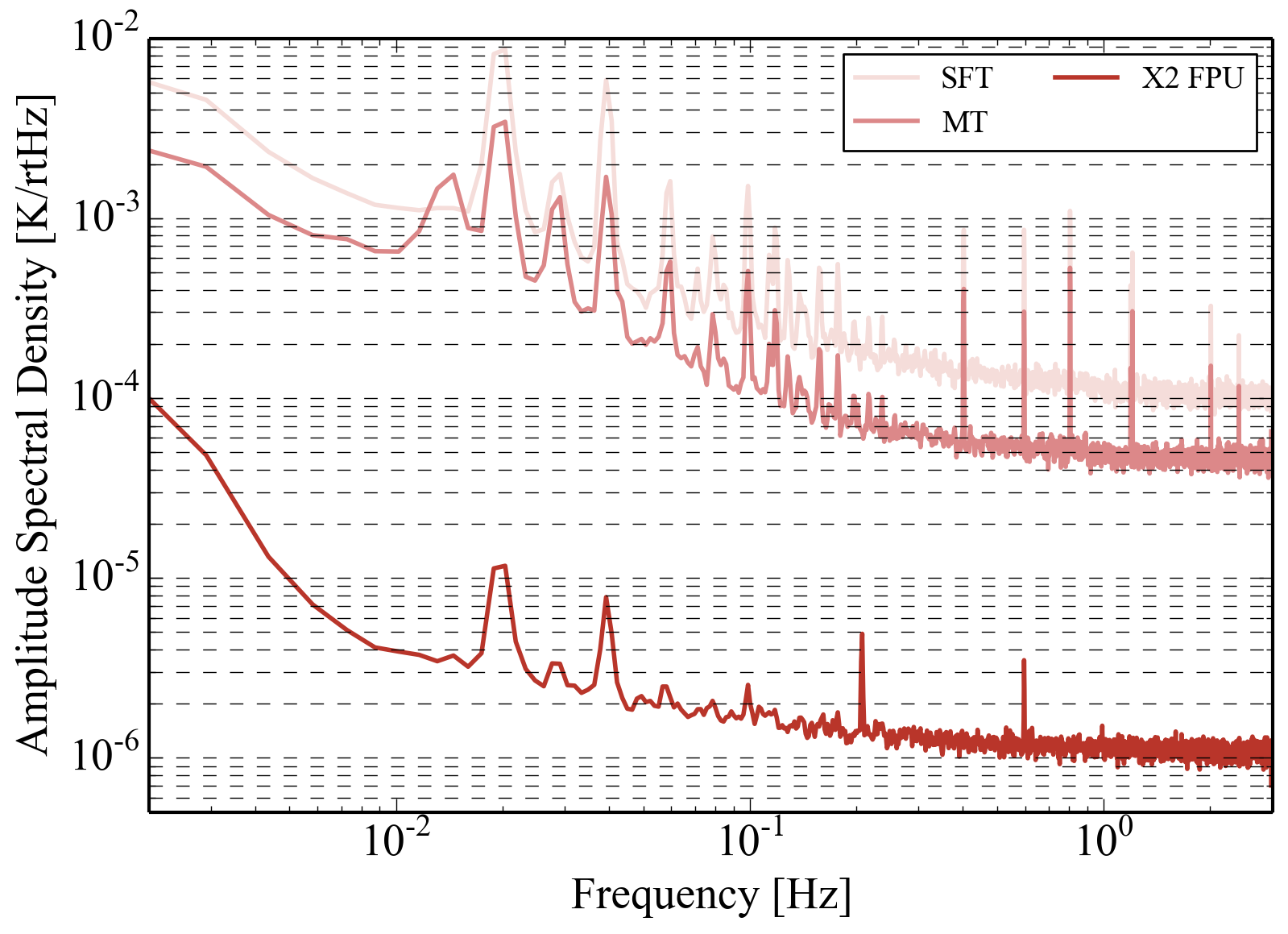}
    \caption[The FPU temperature ASD]{The temperature spectral density as measured by MT and SFT thermometers as well as a Cernox thermometer mounted on the edge of the X2 focal plane. The data are extracted from a five hour period during which all six telescopes were observing the sky. Scan synchronous signal is apparent at 20~mHz and harmonics thereof.}
\label{fig:combo_asd}
\end{center}
\end{figure}

Throughout flight, the half-wave plates were stepped every 12 sidereal
hours and the six adsorption refrigerators were cycled at 8--24 hour
intervals. Figure \ref{fig:fpu} shows the temperature of the six
\spider focal planes over the duration of the flight. Low-amplitude
temperature drifts between fridge cycles are clear, but overall the
focal planes were relatively stable. We note that the focal plane
temperatures for all but the two telescopes mounted in the lowest
ports begins to rise when the main tank liquid level has dropped below
30\% of its full charge. Thankfully, this has a negligible effect on
the detector sensitivity. On average, a focal plane spent 14.9\% of
the flight duration above 350~mK due to scheduled fridge cycles. 
Figure \ref{fig:combo_asd} shows the temperature spectral density as measured by
MT and SFT thermometers (Silicon Diode DT-470) as well as a 
Cernox thermometer mounted on the edge of the X2 focal plane.\footnote{\textit{Lake Shore Cryotronics, Inc.}, Westerville, OH.} The data, derived from a five  
hour period during which all telescopes were operational, provide an 
upper limit for focal plane temperature fluctuations.

Based on integrated flow, we estimate that the final cryostat top-off
resulted in a net charge of 1100~L and that the payload was launched
with approximately 1050~L of liquid helium. During this flight, the
payload successfully observed 10\% of the sky for 16~days.

\section{Conclusions}
\label{sec:conc}

In this paper we have reviewed the design of the \spider balloon-borne
cryostat and highlighted primary results of ground and flight performance. 
Despite significant difference between modeled and realized behavior (see 
Sections \ref{sec:thermal_model} and \ref{sec:cryobe}), the cryostat behaved
adequately on the ground and during flight.  We have demonstrated the suitability of this cryogenic design for future long-duration ballooning flights.

The recovery of the \spider payload is scheduled for the 2015--2016
Antarctic summer. Another flight cryostat, intended for a 
subsequent flight of the \spider experiment, is currently under construction. 

\section{Acknowledgements}

We would like to acknowledge the valuable contributions of Robert
Levenduski, Edward Riedel and Larry Kaylor of Redstone Aerospace to
the design and fabrication of the \spider flight cryostat.

 \spider is supported in the U.S. by National Aeronautics and Space
Administration under Grant No.\ NNX07-AL64G and NNX12AE95G issued
through the Science Mission Directorate, with support for ASR from
NESSF NNX-10AM55H, and by the National Science Foundation \linebreak through
PLR-1043515. Logistical support for the Antarctic deployment and
operations was provided by the NSF through the U.S. Antarctic
Program. The collaboration is grateful for the generous support of the
David and Lucile Packard Foundation, which has been crucial to the
success of the project.

Support in Canada is provided by the National Sciences and Engineering
Council and the Canadian Space Agency.

\bibliography{references}
\bibliographystyle{unsrt85}

\end{document}